\documentclass[12pt,a4paper]{article}

\usepackage{authblk}
\usepackage{amsmath,amssymb}
\usepackage{graphicx,psfrag,color}
\usepackage{amsfonts}
\usepackage{epsfig}
\usepackage{times,pstricks,pst-node}
\usepackage{pstricks,pst-node,pst-text,pst-3d}
\usepackage{caption}

\renewcommand\footnotemark{}

\newcommand{\im}{\mathrm i}

\newcommand{\tr}{\operatorname{Tr}}

\newcommand{\diag}{\operatorname{diag}}
\newcommand{\ket}[1]{\left|#1\right\rangle}      % Ket-Zustand
\newcommand{\braked}[1]{\left\langle #1\right\rangle}      % Ket-Zustand
\newcommand{\bra}[1]{\left\langle #1\right|}     % Bra-Zustand
\newcommand{\eq}{\begin{equation}}
\newcommand{\en}{\end{equation}}
\newcommand{\bear}{\begin{eqnarray}}
\newcommand{\ear}{\end{eqnarray}}

\title{The entropy of the six-vertex model with variety of different boundary conditions}
\author{T.S. Tavares$^{(2)}$, \ G.A.P. Ribeiro$^{(1,2)}$ \ and V.E. Korepin$ ^{(1)}$ \footnote{{\bf E-mails:} tavares@df.ufscar.br;  pavan@df.ufscar.br; korepin@gmail.com}}
\affil{$^{(1)}$ C.N. Yang Institute for Theoretical Physics, \\
State University of New York at Stony Brook, NY 11794, USA \\
$^{(2)}$ Departamento de F\'{i}sica, Universidade Federal de S\~ao Carlos \\ S\~ao Carlos, SP 13565-905, Brazil}
\begin{document}
\maketitle 
\thispagestyle{empty}
\begin{abstract}
We study the dependence of entropy [per lattice site] of six-vertex model on
boundary conditions. We start with lattices of finite size and then proceed
to thermodynamic limit.
We argue that the six-vertex model with periodic, anti-periodic and mixed
boundary conditions produce the same free-energy in the thermodynamic  limit.
We have found fixed boundary conditions such that the entropy varies continously 
from zero to its value for periodic boundary condition.
We have also shown that the physical quantities of the six-vertex model at the 
isotropic point does not change in the case of singular 
toroidal boundary.
\end{abstract}

%\centerline{PACS: 75.10.Pq; 02.30.Ik;}
%\centerline{Keywords: six-vertex model, thermodynamic limit}

\pagestyle{plain}

%\tableofcontents

\section{Introduction}

The six-vertex model has been extensively studied over the years \cite{BAXTER,BOOK}. It was firstly solved under the assumption of periodic boundary conditions\cite{LIEB}. Later on, the equivalence of the six-vertex model with periodic and free boundary conditions was proven by means of the weak-graph expansion \cite{WU}.

Nevertheless, it was also noted in \cite{WU} that the free-energy of the six-vertex model cannot be independent of boundary conditions. Moreover the six-vertex model was again studied in the thermodynamic limit with special free boundaries \cite{OWCZAREK}, anti-periodic boundaries\cite{BATCHELOR}, domain wall boundary \cite{KOREPIN2000,ZINNJUSTIN} and recently the case of domain wall and reflecting end boundary conditions was also considered \cite{RIBEIRO}. On the latter cases of fixed boundary conditions, the free-energy and therefore the entropy was found to differ from the case of periodic boundary condition.

In order to further investigate the dependence of the physical quantities, e.g entropy, of the six-vertex model constrained by different boundary conditions, we started investigating the relation among free boundary, periodic, anti-periodic and mixture of periodic and anti-periodic. Besides that, we have also considered the case of fixed boundary conditions. Our goal is to argue that the entropy of all mixture of periodic and anti-periodic boundaries agrees with the periodic case and that there are different instances of fixed boundary in which the infinite temperature entropy might agrees and disagrees with the periodic or domain wall boundary conditions. 

In particular we introduce what we called N\'eel boundary conditions. This case does not seems to be exactly solvable due to the boundary conditions, however we obtained some results which shows that the entropy is the same as the case of periodic boundary condition. We have also found that this boundary condition is connected with some recent generalization of alternating sign matrix\cite{BRUALDI}.

The outline of the article is as follows. In section \ref{sixvertex}, we describe the six-vertex model and its boundaries conditions. In section \ref{mixed}, we discuss the case of periodic, anti-periodic and mixed boundary conditions. The case of fixed boundary conditions is treated in section \ref{fixed}. In section \ref{6vsing}, we have also addressed to the case of singular boundary conditions at the isotropic point. In appendix A, 
we provide some results for the eight-vertex model for completeness. In appendix B, we discuss the Bethe ansatz solution for $\Delta>1$ and in the appendix C we discuss the case of separated inversions. Our conclusions are given in section \ref{CONCLUSION}.

\section{The six-vertex model}\label{sixvertex}

In this section, we introduce the six-vertex model and its partition function with general free boundary conditions.

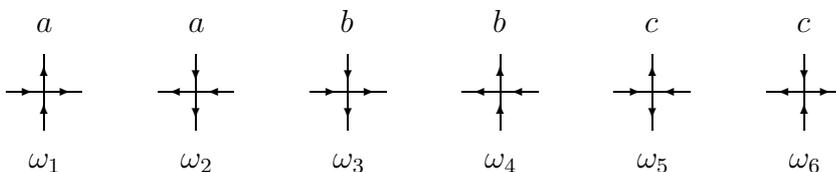
\begin{figure}[h]
\unitlength=1.0mm
\begin{center}
\begin{picture}(125,20)
\put(4,13){$a$}
\put(5,0){\line(0,1){10}}
\put(5,7.5){\vector(0,1){1}}
\put(5,2.5){\vector(0,1){1}}
\put(0,5){\line(1,0){10}}
\put(2.5,5){\vector(1,0){1}}
\put(7.5,5){\vector(1,0){1}}
\put(3,-5){$\omega_1$}
\put(24,13){$a$}
\put(25,0){\line(0,1){10}}
\put(25,7.5){\vector(0,-1){1}}
\put(25,2.5){\vector(0,-1){1}}
\put(20,5){\line(1,0){10}}
\put(22.5,5){\vector(-1,0){1}}
\put(27.5,5){\vector(-1,0){1}}
\put(23,-5){$\omega_2$}
\put(44,13){$b$}
\put(45,0){\line(0,1){10}}
\put(45,2.5){\vector(0,-1){1}}
\put(45,7.5){\vector(0,-1){1}}
\put(40,5){\line(1,0){10}}
\put(42.5,5){\vector(1,0){1}}
\put(47.5,5){\vector(1,0){1}}
\put(43,-5){$\omega_3$}
\put(64,13){$b$}
\put(65,0){\line(0,1){10}}
\put(65,2.5){\vector(0,1){1}}
\put(65,7.5){\vector(0,1){1}}
\put(60,5){\line(1,0){10}}
\put(62.5,5){\vector(-1,0){1}}
\put(67.5,5){\vector(-1,0){1}}
\put(63,-5){$\omega_4$}

\put(84,13){$c$}
\put(85,0){\line(0,1){10}}
\put(85,7.5){\vector(0,1){1}}
\put(85,2.5){\vector(0,-1){1}}
\put(80,5){\line(1,0){10}}
\put(82.5,5){\vector(1,0){1}}
\put(87.5,5){\vector(-1,0){1}}
\put(83,-5){$\omega_5$}
\put(104,13){$c$}
\put(105,0){\line(0,1){10}}
\put(105,7.5){\vector(0,-1){1}}
\put(105,2.5){\vector(0,1){1}}
\put(100,5){\line(1,0){10}}
\put(102.5,5){\vector(-1,0){1}}
\put(107.5,5){\vector(1,0){1}}
\put(103,-5){$\omega_6$}
\end{picture}
\caption{The Boltzmann weights of the six-vertex model.}
\label{6vert}
\end{center}
\end{figure}

The partition function of the statistical model is a sum of all configurations ($\varepsilon$),
\eq
Z=\sum_{\langle \varepsilon \rangle} \prod_{i=1}^N \prod_{j=1}^L \omega^{(i,j)}_{\varepsilon},
\label{partitionZ}
\en
which defines a complicated combinatorial problem. The weight $\omega^{(i,j)}_{\varepsilon}$ can assume the values $a(\lambda), b(\lambda)$ and $c(\lambda)$, which are associated to the different vertices configurations of the six-vertex model (see Figure \ref{6vert}).

These Boltzmann weights are the matrix elements of the so called $R$-matrix \cite{BAXTER,BOOK}, which reads
\eq
R(\lambda)=\left(\begin{array}{cccc}
       a(\lambda) & 0 & 0 & 0 \\
       0 & c(\lambda)& b(\lambda) & 0  \\
       0 & b(\lambda) & c(\lambda) & 0 \\
       0 & 0 & 0 & a(\lambda)
\end{array} \right).
\en

One usually requires that $R$-matrix is a solution of the Yang-Baxter equation,
\eq
R_{12}(\lambda-\mu)R_{23}(\lambda) R_{12}(\mu) =R_{23}(\mu)R_{12}(\lambda) R_{23}(\lambda-\mu),
\label{yangbaxter}
\en
which constraints the Boltzmann weights of the six-vertex such that,
\eq
\Delta=\frac{a^2+b^2-c^2}{2 a b},
\en
for any value of the spectral parameter. On the one hand this impose several constraints on the problem, on the other hand it allows for exact results for certain boundary conditions.

The associativity of Yang-Baxter equation gives rise to the Yang-Baxter algebra \cite{FADDEEV}
\eq
R(\lambda-\mu) \left({\cal T}_{\cal A}(\lambda) \otimes {\cal T}_{\cal A}(\mu)\right) = \left({\cal T}_{\cal A}(\mu) \otimes {\cal T}_{\cal A}(\lambda)\right) R(\lambda-\mu),
\label{fundrel}
\en
where ${\cal T}_{\cal A}(\lambda)={\cal L}_{{\cal A}L}(\lambda-\mu_L)\cdots {\cal L}_{{\cal A}1}(\lambda-\mu_1)$ is the monodromy matrix, ${\cal L}_{12}(\lambda)=P_{12}R_{12}(\lambda)$, $P_{12}$ is the permutation operator and $\cal A$ denotes the auxiliary space along the horizontal direction.

The Yang-Baxter algebra is invariant under transformation of the monodromy matrix \cite{DEVEGA}, such that ${\cal T}_{\cal A}(\lambda) \rightarrow {\cal G}_{\cal A} {\cal T}_{\cal A}(\lambda)$ provided that
\eq
[R(\lambda-\mu),{\cal G}\otimes {\cal G}]=0.
\label{invariance}
\en

For general values of $\Delta$ ($\Delta\neq 1$), this condition implies that the matrix ${\cal G}$ is diagonal or anti-diagonal matrix\cite{DEVEGA}, which reads
\eq
{\cal G}^{(0)}=\left(\begin{array}{cc}
			    1 & 0 \\
			    0 & \alpha_1
                           \end{array}\right),
\qquad
{\cal G}^{(1)}=\left(\begin{array}{cc}
			    0 & 1 \\
			    \alpha_2 & 0
                           \end{array}\right),
\label{discrete}
\en
where we are going to restrict ourselves to the case $\alpha_1=\alpha_2=1$, which give us the relevant matrices to study the free boundary case. The case $\Delta=1$ has $SU(2)$ symmetry, which implies that ${\cal G}$ can be any $2\times 2$ matrix \cite{DEVEGA}.

The equations (\ref{fundrel}-\ref{invariance}) provide the commutativity property of the transfer matrices $T^{(i)}(\lambda)=\tr_{\cal A}\left[ {\cal G}_{\cal A}^{(i)}{\cal T}_{\cal A}(\lambda) \right]$,
\eq
[T^{(i)}(\lambda), T^{(i)}(\mu)]=0, \qquad \forall \lambda,\mu \qquad i=0,1.
\en
However, it is worth to note that $T^{(0)}(\lambda)$ does not commute with $T^{(1)}(\lambda)$.

After a convenient representation of the monodromy matrix
\eq
{\cal T}_{\cal A}(\lambda)=\left(\begin{array}{cc}
                            A(\lambda) & B(\lambda) \\
                            C(\lambda) & D(\lambda)
                           \end{array}\right),
\en
we clearly see that the transfer matrix are simply given by
\bear
T^{(0)}(\lambda)&=& A(\lambda) + D(\lambda), \\
T^{(1)}(\lambda)&=& B(\lambda) + C(\lambda),
\ear
where these are the transfer matrices with periodic and anti-periodic boundary conditions respectively.

The transfer matrices $T^{(i)}(\lambda)$, when multiplied successively, builds up the partition function of a $N\times L$ bidimensional classical vertex model with some boundary conditions.

In the case of free boundary condition, we have that the partition function (\ref{partitionZ}) can be written as
\bear
Z_{free}&=&\sum_{\phi_k,\theta_j=0,1}\tr_V{\left[\bigotimes_{k=1}^L {\cal G}_{V_k}^{(\phi_k)} \prod_{j=1}^N T^{(\theta_j)}(\lambda_j)\right]},
\label{partfree}
\ear
where in each external bond one could have all possible configurations of incoming and outgoing arrows. The matrix ${\cal G}_V^{(\phi)}$ is taken from (\ref{discrete}) and it stands for the periodic or anti-periodic boundary on the vertical direction $V$. Its product $\bigotimes_{k=1}^L {\cal G}_{V_k}^{(\phi_k)}$ produce all possible configurations of incoming and outgoing arrows along the vertical direction, where $\phi_k=0,1$ for $k=1,\dots,L$. On the other hand, the product of transfer matrices for $\theta_j=0,1$ for $j=1,\dots,N$ produce all possible products of $T^{(0)}(\lambda)$ and $T^{(1)}(\lambda)$ and therefore all possible configurations of incoming and outgoing arrows along the horizontal direction.

\section{The six-vertex model with mixed boundary conditions}\label{mixed}

In the homogeneous case ($\lambda_j=\lambda,\mu_k=\mu=0$) the expression (\ref{partfree}) is simply given by
\bear
Z_{free}&=& \tr_V{\left[\bigotimes_{k=1}^L \left(\begin{array}{cc}
                            1 & 1 \\
                            1 & 1
                           \end{array}\right)_k (A(\lambda)+D(\lambda)+B(\lambda)+C(\lambda))^N\right]}.
\ear

Naturally, the case of free boundary contains as special cases other more constrained boundary conditions. The case of periodic boundary condition ($P$) along both directions consist of the term $\phi_k=\theta_j=0, \forall j,k$  in (\ref{partfree}). This term can be simply written in the homogeneous case as
\bear
Z_{PP}&=&\tr_V{\left[(T^{(0)}(\lambda))^N\right]}= \tr_V{\left[(A(\lambda)+D(\lambda))^N\right]},
\label{PP}
\ear
and was studied in \cite{LIEB}. On the other hand, the case of anti-periodic ($A$) boundary along the horizontal and periodic along the vertical direction $(\theta_j=1,\phi_k=0,  \forall j,k)$ can be written in the homogeneous case as,
\bear
Z_{AP}&=&\tr_V{\left[(T^{(1)}(\lambda))^N\right]}= \tr_V{\left[(B(\lambda)+C(\lambda))^N\right]}.
\label{AP}
\ear
This case was considered in \cite{BATCHELOR} in the anti-ferroelectric regime ($\Delta<-1$).

Besides the immediate cases of periodic and anti-periodic boundary $Z_{PA}$ as well the anti-periodic boundary in both direction $Z_{AA}$, one has a number of additional non-trivial mixed boundaries. One can organize this in matrix such that,
\bear
M_{N,L}=\left(\begin{array}{cccccc}
Z_{1,1} &  Z_{1,2}  & \cdots & Z_{1,2^L} \\
Z_{2,1} &  Z_{2,2}  & \cdots & Z_{2,2^L} \\
\vdots & \vdots &  \ddots  & \vdots \\
Z_{2^N,1} &  Z_{2^N,2} & \cdots & Z_{2^N,2^L}
\end{array}\right), \qquad Z_{free}=\sum_{j=1}^{2^N}\sum_{k=1}^{2^L} Z_{j,k}.
\label{countmixed}
\ear
We have that $Z_{j,k}$ is the partition function of lattice $N\times L$ with ${(j-1)}_{10}={\{\theta_N,\dots,\theta_1\}}_{2}$ and ${(k-1)}_{10}={\{\phi_L,\dots,\phi_1\}}_{2}$ defining local periodic or anti-periodic closing along the horizontal and vertical directions, respectively. It is to be understood in the above notation that the binary representation of $j-1$ gives us the vector $\{\phi_L,\dots,\phi_1\}$ and similarly for $k-1$ and $\{\theta_N,\dots,\theta_1\}$. More explicitly, for the integers $j$ and $k$ we have that $j-1=\theta_1 2^0 + \theta_2 2^1 +\dots +\theta_N 2^{N-1}$ and $k-1=\phi_1 2^0 + \phi_2 2^1 +\dots +\phi_L 2^{L-1}$; hence
\eq
Z_{j,k}=\tr_V{\left[\bigotimes_{m=1}^L {\cal G}_{V_m}^{(\phi_m)} \prod_{n=1}^N T^{(\theta_n)}(\lambda) \right]}, j=1,\dots,2^N, k=1,\dots,2^L,
\label{generalpart}
\en
where for a given $j$ and $k$ on the left hand side we have a set of $\phi_m$ and $\theta_n$ on the right hand side. We also have that $Z_{1,1}=Z_{PP}$, $Z_{2^N,1}=Z_{AP}$, $Z_{1,2^L}=Z_{PA}$ and $Z_{2^N,2^L}=Z_{AA}$. 

In order to illustrate, let us consider the instance of $L=N=2$. On the one hand, the first column ($k=1$) of the matrix (\ref{countmixed}), which means $(0)_{10}={\{\phi_2=0,\phi_1=0\}}_{2}$,  represents the case of periodic boundary conditions along the vertical direction. This is because ${\cal G}_{V_m}^{(0)}$ is the identity matrix (\ref{discrete}). On the other hand, we have the following cases along the horizontal direction:

i) periodic boundary at $j=1$ which again implies $\theta_n=0$ for $n=1,2$ and therefore $$Z_{1,1}=\tr_V{\left[(T^{(0)}(\lambda))^2\right]};$$

ii) periodic and anti-periodic at $j=2$ ($(1)_{10}={\{\theta_2=0,\theta_1=1\}}_{2}$) $$Z_{2,1}=\tr_V{\left[T^{(0)}(\lambda)T^{(1)}(\lambda)\right]};$$

iii) anti-periodic and periodic at $j=3$ ($(2)_{10}={\{\theta_2=1,\theta_1=0\}}_{2}$) $$Z_{3,1}=\tr_V{\left[T^{(1)}(\lambda)T^{(0)}(\lambda)\right]};$$

iv) finally the anti-periodic case at $j=4$ ($(3)_{10}={\{\theta_2=1,\theta_1=1\}}_{2}$) $$Z_{4,1}=\tr_V{\left[(T^{(1)}(\lambda))^2\right]}.$$

For more general matrix elements, one should also take in account the ${\cal G}_V^{(\phi)}$ boundary matrices ($k>1$) which will reverse or not the edge configurations along the vertical direction depending on $k$ value. Although we have explicitly given $Z_{2,1}$ and $Z_{3,1}$ in the above example, in this case they are actually vanishing due to the property of the arrow flux along the boundary.

More specifically, we have trivial cases which results in a vanishing partition function. In general, the non-trivial cases should fulfill the following rule: Let $\Phi=\sum_{m=1}^L \phi_m$ be the number of vertical anti-periodic closings and similarly $\Theta=\sum_{n=1}^N \theta_n$ the number of horizontal anti-periodic closings, then $\mbox{mod}\left[\Phi-\Theta,2\right]$ must be zero. This condition comes from the parity of the flux of arrow on the boundary.

In addition to that, the partition functions $Z_{j,k}$ is related under $\pi/2$ rotation of the lattice with $Z_{k,j}$, which reads
\eq
Z_{j,k}:=Z_{j,k}^{N \times L}=Z_{k,j}^{L \times N}.
\label{symmetry}
\en

All the above features can be seen at the infinite temperature point. By tuning the Boltzmann weights such that $a=b=c=1$ ($\Delta=1/2$), we are dealing with the infinite temperature case. Therefore, the partition function is just counting the number of equally likely physical states $Z=\Omega$. We have obtained the number of physical states ($\Omega$) for small sizes (up to $N=L=5$). The cases $N=L=2,3$ are given below,
\bear
M_{2,2}=\left(\begin{array}{cc|cc}
18 &  0 & 0 & 8 \\
0 &  10 & 10 & 0 \\
\hline
0 &  10 & 10 & 0 \\
8 &  0 & 0 & 8 \\
\end{array}\right),
M_{3,3}=\left(\begin{array}{cccc|cccc}
148 &  0 & 0 & 84 & 0 & 84 & 84 & 0 \\
0 &  94 & 84 & 0 & 94 & 0 & 0  & 72 \\
0 &  84 & 80 & 0 & 84 & 0 & 0  & 72 \\
84 &  0 & 0 & 74 & 0 & 72 & 74 & 0 \\
\hline
0 & 94 & 84 & 0 & 94 & 0 & 0 & 72 \\
84& 0 &  0 & 72 & 0 & 76 & 72 & 0 \\
84& 0 &  0 & 74 & 0 & 72 & 74 & 0 \\
0 & 72 & 72 & 0 & 72 & 0 & 0 & 68
\end{array}\right).
\label{countmixedexp}
\ear
One can see that the case of periodic boundary along both directions allows for the largest number of configurations. As a result, we can show that the entropy of six-vertex model at infinite temperature with periodic boundary agrees with the case of free boundary. This is done by noting that the number of configurations of free boundary ($\Omega_{free}$), which is equal to the sum of all matrix elements of $M_{N,L}$ at $\Delta=1/2$, is larger than the number of configurations of the periodic case $\Omega_{PP}=Z_{1,1}$. Besides that, we estimate an upper bound for $\Omega_{free}$ assuming that all the $2^{L+N-1}$ non-trivial boundaries (the non-vanishing matrix elements of $M_{N,L}$) are equal to the maximal value $\Omega_{PP}$. This implies that,
\eq
\Omega_{PP} \leq \Omega_{free} \leq 2^{L+N-1} \Omega_{PP}.
\label{ineq}
\en
In order to take the thermodynamic limit, we raise (\ref{ineq}) to the power $1/(N L)$, take the logarithm and at last take the infinite size limit. This shows that the infinite temperature entropy of the free boundary case agrees with the entropy of the boundary which allows for the largest number of configurations, therefore periodic boundary conditions ($S_{PBC}=S_{free}$).

It is worth to note that this relation among six-vertex model with periodic and free boundary conditions at arbitrary temperatures was proven for even lattice sizes by means of the weak-graph expansion \cite{WU}. Although our reasoning is limited to infinite temperatures at this point, there is no restriction on the lattice sizes. In the next section, we are going to extend this result to different temperatures.

Besides that, the above imbalance in the number of states among the different mixed boundary (\ref{countmixedexp}) is due to the ice-rule. In the case of the eight-vertex model (see appendix A), this imbalance does not exist due to the additional allowed vertex configurations.

\subsection{Homogeneous boundary conditions}

The cases $Z_{PP}, Z_{AP}, Z_{PA}$ and $Z_{AA}$ are build up by the product of $T^{(0)}(\lambda)$ or $T^{(1)}(\lambda)$ only. This implies that we have products of commuting operators, which by its turn are integrable transfer matrices. Therefore, the analysis of these cases are much simpler than the more general cases $Z_{j,k}$.

In order to analyze  all the above cases, we will exploit some discrete symmetries of the transfer matrices $T^{(i)}(\lambda)$.

We can define the reflection operator as $\Pi^x = \bigotimes_{m=1}^L \sigma^x_m$ and the parity operator as $\Pi^z= \bigotimes_{m=1}^L \sigma^z_m$, where $\sigma^{x,y,z}$ denotes the usual Pauli matrices. These are special cases of the discrete symmetries (\ref{discrete}), which implies the following commutation relations
\begin{align}
&&\left[T^{(0)}(\lambda),\Pi^{x}\right]&=\left[T^{(0)}(\lambda),\Pi^{z}\right]=0, \label{symT0}\\
&&\left[T^{(1)}(\lambda),\Pi^x\right]&=\left[T^{(1)}(\lambda),\Pi^{z}\right]_{+}=0,  \label{symT1} \\
&&\Pi^x\Pi^z&={(-1)}^{L} \Pi^z \Pi^x, \label{symPXPZ}
\end{align}
which comes from the invariance (\ref{invariance}) and the last relation is a direct a byproduct of algebraic properties of the Pauli matrices.

\subsubsection{Completely periodic boundary condition}

The case of periodic boundary conditions $Z_{PP}$ was completely solved long ago by standard Bethe ansatz\cite{LIEB}. The partition function $Z_{PP}$ can be written as
\bear
Z_{PP}&=& \sum_{j=1}^{2^{L}} (\Lambda_{j}^{(0)}(\lambda))^N,
\ear
where $\Lambda_j^{(0)}(\lambda)$ are the transfer matrix eigenvalues\cite{LIEB}. Therefore it is immediate that only the largest eigenvalue $\Lambda_{max}^{(0)}(\lambda)$ contributes for the free-energy in the thermodynamic limit, such that
\eq
F_{PP}=-\frac{1}{\beta} \lim_{L,N\rightarrow \infty} \frac{1}{L N}\ln{(\Lambda_{max}^{(0)}(\lambda))^N},
\label{FPP}
\en
where $\beta$ is the inverse of temperature. By its turn, the largest eigenvalue of $T^{(0)}(\lambda)$ and therefore the free-energy in the thermodynamic limit is known for all values of $\Delta$ \cite{LIEB,BAXTER}. The entropy at infinite temperature was given as\cite{LIEB}
\eq
S_{PBC}=\frac{1}{2} \ln\left(\frac{4}{3}\right)^3\approx 0.431523.
\label{entrPBC}
\en

\subsubsection{Anti-periodic boundary along the horizontal direction}

On the other hand, as the case  $Z_{AP}$ lacks of the arrow conservation flux from row to row, it was solved much later by means of the $T-Q$ approach \cite{BATCHELOR}. It was shown in \cite{BATCHELOR} for even $L$ and $\Delta<-1$ that the partition function $Z_{AP}$ produce the same free-energy as the case of periodic boundary condition \cite{LIEB}.

In order to better understand this result, it is convenient to exploit the anti-commutation rule between $\Pi^z$ and $T^{(1)}(\lambda)$. In doing so, we obtain that in the case that $\ket{\Psi}$ is an eigenvector of $T^{(1)}(\lambda)$ with eigenvalue $\Lambda(\lambda)$, one also has that $\Pi^z \ket{\Psi}$ is an eigenvector with eigenvalue $-\Lambda(\lambda)$. In other words, all the eigenvalues of $T^{(1)}(\lambda)$ are at least double-degenerate in modulus and we may write,
\eq
T^{(1)}(\lambda)=\sum_{j=1}^{2^{L-1}} \Lambda_{j}^{(1)}(\lambda) \ket{\Psi_j^{(1)}}\bra{\Psi_j^{(1)}}-\Lambda_{j}^{(1)}(\lambda) \Pi^z\ket{\Psi_j^{(1)}}\bra{\Psi_j^{(1)}}\Pi^z,
\label{T1}
\en
where we can make any choice of the first $2^{L-1}$ states by selecting one state between each pair $\{\Lambda_j^{(1)}(\lambda)$, $-\Lambda_j^{(1)}(\lambda)\}$.

Therefore the partition function results in,
\eq
Z_{AP}=\tr_{V}\left[ (T^{(1)}(\lambda))^N\right]=
\begin{cases}
0 & \text{odd $N$}, \\
\displaystyle 2 \sum_{j=1}^{2^{L-1}} (\Lambda_{j}^{(1)}(\lambda))^N &\text{even $N$},
\end{cases}
\label{evenPA}
\en
which again implies that only the largest eigenvalue $\Lambda_{max}^{(1)}(\lambda)$ contributes for the free-energy in the thermodynamic limit,
\eq
F_{AP}=-\frac{1}{\beta} \lim_{L,N\rightarrow \infty} \frac{1}{L N}\ln{(\Lambda_{max}^{(1)}(\lambda))^N}.
\en

The fact that $F_{AP}=F_{PP}$ establishes the following relation between the maximal transfer matrix eigenvalues,
\eq
\lim_{L\rightarrow \infty} \frac{1}{L}\ln{\Lambda_{max}^{(1)}(\lambda)} = \lim_{L\rightarrow \infty} \frac{1}{L}\ln{\Lambda_{max}^{(0)}(\lambda)},
\label{relL0L1}
\en
at least for $\Delta<-1$ thanks to the result in \cite{BATCHELOR}.

\subsubsection{Completely anti-periodic boundary condition}

Let us now address to the case of anti-periodic boundary conditions along both directions $Z_{AA}$. In this case, due to the relation (\ref{symPXPZ}), we have to distinguish between odd and even values of $L$.

For $L$ odd, the anti-commutation between $\Pi^x$ and $\Pi^z$ also implies that the $\Pi^x$ eigenvalues for $\ket{\Psi}$ and $\Pi^z \ket{\Psi}$ are opposite. In this case, we choose the first $2^{L-1}$ states to have $\Pi^x$ eigenvalues $+1$, which results in
\eq
\Pi^x (T^{(1)}(\lambda))^N=\sum_{j=1}^{2^{L-1}} (\Lambda_{j}^{(1)}(\lambda))^N \bigg[ \ket{\Psi_j^{(1)}}\bra{\Psi_j^{(1)}}-{(-1)}^N \Pi^z\ket{\Psi_j^{(1)}}\bra{\Psi_j^{(1)}}\Pi^z \bigg].
\en
Taking the trace along the vertical direction, we obtain
\eq
Z_{AA}=Z_{2^N,2^L}=\tr_{V}\left[\Pi^x (T^{(1)}(\lambda))^N\right]=
\begin{cases}
\displaystyle 2 \sum_{j=1}^{2^{L-1}} (\Lambda_{j}^{(1)}(\lambda))^N &\text{for $N$ odd},\\
0 & \text{for $N$ even},
\end{cases}
\label{oddaa}
\en
which agrees with the matrix element values $Z_{2^N,2^L}$ of (\ref{countmixed}), as verified for $N,L \leq 5$, $L$ odd.

On the other hand, the operators $\Pi^x$ and $\Pi^z$ do commute for $L$ even, which implies the eigenvectors $\ket{\Psi}$ and $\Pi^z \ket{\Psi}$ have simultaneously the same $\Pi^x$ eigenvalues. For this reason, we choose an ordering such that the first $2^{L-1}$ eigenvectors alternates between $\Pi^x$ eigenvalues $+1$ and $-1$. Therefore
\eq
\Pi^x (T^{(1)}(\lambda))^{N}=\sum_{j=1}^{2^{L-1}} {(-1)}^{j-1} (\Lambda_{j}^{(1)}(\lambda))^N\bigg[ \ket{\Psi_j^{(1)}}\bra{\Psi_j^{(1)}}+{(-1)}^N \Pi^z\ket{\Psi_j^{(1)}}\bra{\Psi_j^{(1)}}\Pi^z\bigg],
\en
which implies
\eq
Z_{AA}=\tr_{V}\left[\Pi^x (T^{(1)}(\lambda))^N\right]=
\begin{cases}
0 &\text{for $N$ odd},\\
\displaystyle 2 \sum_{j=1}^{2^{L-1}} {(-1)}^{j-1} (\Lambda_{j}^{(1)}(\lambda))^N  & \text{for $N$ even}.
\end{cases}
\label{evenaa}
\en
This again agrees with matrix element $Z_{2^N,2^L}$ of (\ref{countmixed}). The selection rule $\mbox{Mod}\left[\Phi-\Theta,2\right]=0$ is naturally fulfilled by the relations (\ref{oddaa}) and (\ref{evenaa}).

These results imply again that only the largest eigenvalue $\Lambda_{\max}^{(1)}(\lambda)$ contributes for the free-energy in the thermodynamic limit,
\eq
F_{AA}=-\frac{1}{\beta} \lim_{L,N\rightarrow \infty} \frac{1}{L N}\ln{(\Lambda_{\max}^{(1)}(\lambda))^N},
\en
which due to (\ref{relL0L1}) agrees with the periodic case at least for $\Delta<-1$.

\subsubsection{Anti-periodic boundary along the vertical direction}

The case $Z_{PA}$ is related to $Z_{AP}$ due to the rotational symmetry (\ref{symmetry}) and therefore it should be related to the periodic case. In the next section we will further comment on this case.

\subsubsection{Largest partition function}

Based on the direct computation of the partition function and on the above results, we can extend our previous reasoning about the agreement between the free-energy for free and periodic boundary conditions at infinite temperature.
In fact, the free-energy of these boundary conditions also agree for every set of positive Boltzmann weights $a,~b,~c$, which holds true for every temperature value. This can be seen by noting that $Z_{PP}, Z_{AP}, Z_{PA}$ and $Z_{AA}$ are the only matrix elements of (\ref{countmixed}) that can become the dominant term as $a,~b,~c$ change.

In general, we have that $Z_{PP}$ has the largest value for $\Delta \ge -1$. However, the situation changes when $\Delta < -1$. In the anti-ferroelectric phase, the largest term changes according to the parity of the lattice size $L$ and $N$ (see Table \ref{table1}).
\begin{table}[tbh]
\begin{center}
\begin{tabular}{|c|c|}
  \hline
   & Largest contribution for $\Delta<-1$ \\ \hline
  $L$ even, $N$ even & $Z_{PP}$ \\ \hline
  $L$ even, $N$ odd & $Z_{PA}$ \\ \hline
  $L$ odd, $N$ even & $Z_{AP}$ \\ \hline
  $L$ odd, $N$ odd & $Z_{AA}$ \\
  \hline
\end{tabular}
\end{center}
\caption{Largest element of $M_{N,L}$ for $\Delta<-1$.}
\label{table1}
\end{table}

This implies that the largest term $Z_{\max}=\max\left[Z_{PP},Z_{PA},Z_{AP},Z_{AA}\right]$, which should satisfies
\eq
Z_{\max} \leq Z_{free} \leq 2^{L+N-1} Z_{\max},
\en
is the one whose free-energy $F_{\max}$ agrees with the free boundary case. Therefore, whenever $Z_{AP}$, $Z_{PA}$ or $Z_{AA}$ is maximal, we have that $F_{AP}=F_{free}$, $F_{PA}=F_{free}$ or $F_{AA}=F_{free}$. On the top of that, the equality (\ref{relL0L1}) imply that all of them should be equal $F_{AP}=F_{PA}=F_{AA}=F_{PP}=F_{free}$ for $\Delta<-1$ and provided that they are allowed by selection rule. Therefore, even in the case where $Z_{PP}$ is not the largest term, we still have $F_{PP}=F_{free}$.

\subsection{Mixed boundary terms}

Now we want to address to the case of mixed boundary terms $Z_{j,k}$. These are generically a mixed product of the transfer matrices $T^{(0)}(\lambda)$ and  $T^{(1)}(\lambda)$. As these transfer matrix do not commute, we loose the integrability of the statistical model. Although we still can diagonalize the transfer matrices exactly, we have now to deal with the projection of one set of eigenvectors onto another. We want to argue whether or not these mixed boundary partition function will lead to the same free-energy as the periodic case. We show that even in this non-integrable case, we can still extract some information about the thermodynamic limit.

\subsubsection{First row}

Although the first and last rows of (\ref{countmixed}) seems to be of mixed type, they are actually a product of transfer matrix of the same kind which still preserves the integrability. However, we are going to look at the first row case $Z_{1,j}$ in order to better understand the structure of the remaining mixed terms.

Therefore, we address to the case of totally periodic boundary condition on the horizontal direction ($\Theta=0$) but quite general combination of periodicity and anti-periodicity on the vertical direction ($\Phi\neq 0$). The equation (\ref{generalpart}) simplifies to
\eq
Z_{1,j}=\tr_V \left[\bigotimes_{m=1}^L{\cal G}_{V_m}^{(\phi_m)} (T^{(0)}(\lambda))^N\right]=\sum_{g} {\left(\Lambda_g^{(0)}(\lambda)\right)}^N f_{L,g}^{\{\phi_m\}} ,
\label{horizontalgeneral}
\en
where $g^{(0)}$ stands for the $g$-th eigenvector of $T^{(0)}(\lambda)$ and the function $f_{L,g}^{\{\phi_m\}}=\bra{g^{(0)}} \prod_{m=1}^L{\cal G}_{V_m}^{(\phi_m)} \ket{g^{(0)}}$. 

We can order these eigenvectors $\ket{g^{(0)}}$ and consequently the eigenvalues $\Lambda_g^{(0)}(\lambda)$ by the eigenvalues of total spin-$z$ component, since $T^{(0)}(\lambda)$ commutes with the total spin operator $S^z$. 
The eigenvalues of $T^{(0)}(\lambda)$ are given by means of Bethe ansatz\cite{BAXTER},
\eq
\Lambda_n(\lambda)= (a(\lambda))^L \prod_{i=1}^n \frac{a(\lambda_i-\lambda)}{b(\lambda_i-\lambda)} + (b(\lambda))^L \prod_{i=1}^n \frac{a(\lambda-\lambda_i)}{b(\lambda-\lambda_i)}, ~~ n=0,\dots,L
\en
in terms of the solution of the Bethe ansatz equation
\eq
{\left[\frac{a(\lambda_i)}{b(\lambda_i)}\right]}^L= \prod_{\stackrel{j=1}{j \neq i}  }^n \frac{a(\lambda_i-\lambda_j)}{b(\lambda_i-\lambda_j)} \frac{b(\lambda_j-\lambda_i)}{a(\lambda_j-\lambda_i)}, \qquad i=1, \dots, n,
\label{Bet}
\en
where the sectors $n=0$ and $n=L$ correspond to the up and down ferromagnetic state and the intermediate value $n=L/2$ ($L$ even) corresponds to the sector with zero total spin.

For the case $\Delta<1$ and $L$ is even, the largest eigenvalue $\Lambda_{\max}^{(0)}(\lambda)$ is non-degenerate and it is found in sector $n=\frac{L}{2}$. On the other hand, the  for the odd $L$ case, the largest eigenvalue is doubly degenerate. In this case, it has one eigenvalue in sector $n=\frac{L-1}{2}$ and other in sector $n=\frac{L+1}{2}$.

Nevertheless in both cases, the eigenvectors of the largest eigenvalues are complicated linear combinations of all canonical basis vectors belonging to that sector. The coefficients of this linear combination are all positive by virtue of Perron-Frobenius theorem, therefore so it is the expectation value in (\ref{horizontalgeneral}). 
As a result, the action of $\prod_{m=1}^L{\cal G}_{V_m}^{(\phi_m)}$ over the eigenvector always generates a vector with at least one positive component in the canonical basis element of that sector whenever $\mbox{Mod}[\Phi,2]=0$.
Hence the coefficient $f_{L,g}^{\{\phi_m\}}$ is an unknown function of $L$, but most importantly it is independent of $N$. This implies that this coefficient simply do not contribute to the free-energy in the thermodynamic limit,
\eq
F_{1,j}=-\frac{1}{\beta} \lim_{L,N \rightarrow \infty} \frac{1}{N L}\ln{\left[(\Lambda_{\max}^{(0)}(\lambda))^N  f_{L,g_{\max}}^{\{\phi_m\}}\right]}=-\lim_{L \rightarrow \infty} \frac{1}{L}\ln{\Lambda_{\max}^{(0)}(\lambda)}.
\label{resultdmenor1}
\en
Here we assume that $f_{L,g_{\max}}^{\{\phi_{m}\}}>\exp(-\delta L)$ for large $L$ and any $\delta>0$, which was verified for finite lattices.

The case $\Delta>1$ is more subtle. In this case, the leading term of the expression (\ref{horizontalgeneral}) changes with the number of inversions $\Phi$, although the largest eigenvalue of the transfer matrix is always in the sectors $n=0$ and $n=L$. This is due to the fact that the coefficient $f_{L,g}^{\{\phi_m\}}$ is non-vanishing in the sectors  $n=0$ and $n=L$ only at $\Phi=0$, which is the periodic boundary case. In order to see that, we write both ${\cal G}^{(0)}$ and ${\cal G}^{(1)}$ in terms of eigenvectors of ${\cal G}^{(1)}$, such that
\eq
\mathcal{G}^{(\phi)}=\sum_{s=0,1} {(-1)}^{s \phi} \ket{s}\bra{s}, \qquad \ket{s}=\frac{1}{\sqrt{2}}\left(\ket{\uparrow}+{(-1)}^{s}\ket{\downarrow}\right).
\en
This implies that
\eq
\prod_{m=1}^L{\cal G}_{V_m}^{(\phi_m)}=\sum_{\langle s \rangle} \exp\left(\pi \im \sum_{j=1}^L s_j \phi_j\right) \ket{s_1,\dots,s_L}\bra{s_1,\dots,s_L},
\en
 and since we know exactly the eigenvectors in sectors $n=0,L$, we can easily represent them in basis generated by $\ket{s_1,\dots,s_L}$,
\bear
\ket{\uparrow,\dots,\uparrow}&=&\frac{1}{2^{\frac{L}{2}}} \sum_{\langle s \rangle } \ket{s_1,\dots,s_L}, \\
\ket{\downarrow,\dots,\downarrow}&=&\frac{1}{2^{\frac{L}{2}}} \sum_{\langle s \rangle } \exp\left(\pi \im \sum_{m=1}^L s_m\right)\ket{s_1,\dots,s_L},
\ear
and therefore we find
\eq
Z_{1,j}= {\left( \Lambda_{\max,n=0}(\lambda)\right)}^{N} \frac{1}{2^{L-1}} \prod_{k=1}^L \left(1+{(-1)}^{\phi_k}\right)~~+~~ \text{terms from other sectors}.
\label{other}
\en
One can see that this expression is reduced to the case of periodic boundary condition when $j=1$ ($\phi_k=0$).

The other terms in (\ref{other}) occurs for $j>1$ ($\Phi>0$). According to the selection rule $\mbox{Mod}[\Phi,2]=0$, the coefficient $f_{L,g}^{\{\phi_m\}}$ vanishes for $\Phi=1$. Therefore, the simplest non-trivial case is $\Phi=2$. In this case, as the coefficient of the largest eigenvalues in the sectors $n=0,L$ still vanishes, we need to proceed to the sectors $n=1,L-1$, which contain the next leading eigenvalue. Once again, we know the eigenvectors exactly because $T^{(0)}(\lambda)$ commutes with translation operator. The eigenvectors of sectors $n=1,L-1$ can be written as
\bear
\ket{\Psi_{m,n=1}^{(0)}} &=&\sum_{k=1}^{L} \frac{{\rm e}^{-\frac{2 \pi \im}{L} (k-1) (m-1)}}{\sqrt{L}} \ket{\downarrow_k}, \\
\ket{\Psi_{m,n=L-1}^{(0)}}&=&\sum_{k=1}^{L} \frac{{\rm e}^{-\frac{2 \pi \im}{L} (k-1) (m-1)}}{\sqrt{L}} \ket{\uparrow_k}, \qquad m=1,\dots,L,
\ear
whose leading term eigenvector occurs for $m=1$. Therefore, we have
\eq
Z_{1,j} =\frac{4}{L} { \left(\Lambda_{\max,n=1}(\lambda)\right)}^N ~~ + ~~\text{negligible terms in the thermodynamic limit},
\en
hence
\eq
F_{1,j}=-\frac{1}{\beta} \lim_{L \rightarrow \infty} \frac{1}{L}\ln{\left(\Lambda_{\max,n=1}(\lambda)\right)}=-\frac{1}{\beta}\lim_{L \rightarrow \infty}\frac{1}{L}  \ln{\left(\Lambda_{\max,n=0}(\lambda)\right)}=F_{PP},
\en
whenever $j$ contains exactly two inversions. The equality from the above limits comes from the fact that the ratio between the two eigenvalues tends to a non-zero constant when $L$ goes to infinity, and this is sufficient the give the same limit.

Quite generally we have the following results for $\Delta>1$. The partition function $Z_{1,j}$ is zero for $\Phi= 2 k-1$. For $\Phi =2 k$, the partition function per site in thermodynamic limit can be computed from the largest eigenvalue in sector $k$, such that
\eq
F_{1,j}=-\frac{1}{\beta} \lim_{L \rightarrow \infty}  \frac{1}{L}\ln{\left(\Lambda_{\max,n=k}(\lambda)\right)}.
\en
This is a consequence of the zero expectation value of $\prod_{m=1}^L{\cal G}_{V_m}^{(\phi_m)}$ for eigenvectors in sectors $n$ such that $n<k$ or $n>L-k$. Specifically when $\Phi=L$, for $L$ even, we have
\eq
F_{PA}=-\frac{1}{\beta} \lim_{L \rightarrow \infty}   \frac{1}{L}\ln{\left(\Lambda_{\max,n=\frac{L}{2}}(\lambda)\right)}.
\label{resultdmaior1}
\en
This result seems to differ from the case of periodic boundary conditions (\ref{FPP}) since for $\Delta>1$ the largest eigenvalue is given $\Lambda_{\max}^{(0)}(\lambda)=\Lambda_{n=0}(\lambda)$ and here the leading contribution comes from largest eigenvalue in the sector $n=L/2$ $\Lambda_{\max,n=\frac{L}{2}}(\lambda)$. However, as noted before the ratio of these eigenvalues goes to a constant value for large $L$. This confirms that the $F_{PA}=F_{PP}$. This can also be proved by means of the analysis of the Bethe ansatz solution for the eigenvalue $\Lambda_{\max,n=\frac{L}{2}}(\lambda)$, as described in the appendix B.

\subsubsection{First column}

Due to the symmetry  (\ref{symmetry}), the first column and the first row should produce the same results. However, their explicit expression are quite different in view of representation (\ref{generalpart}). Nevertheless, it is still interesting to look at $Z_{j,1}$ in order to understand another piece of the structure of the general terms $Z_{j,k}$.

The partition function with periodic boundary condition on the vertical direction and mixed boundary conditions along the horizontal can be written as,
\eq
Z_{j,1}=\tr\left[\prod_{m=1}^N T^{(\theta_m)}(\lambda)\right],
\label{verticalgeneral}
\en
where the first and last terms $Z_{PP}$ and $Z_{AP}$ were already discussed in the previous sections.

In view of symmetry property (\ref{symmetry}), the result (\ref{resultdmenor1}) for $\Delta<1$, the result (\ref{resultdmaior1}) for $\Delta>1$, we can extend (\ref{relL0L1}) as
\eq
\lim_{L \rightarrow \infty} \frac{1}{L}\ln{\Lambda^{(0)}_{\max,n=\frac{L}{2}}(\lambda)}= \lim_{L \rightarrow \infty}  \frac{1}{L}\ln{\Lambda_{\max}^{(1)}(\lambda)} ,
\en
for all values of $\Delta$. Note that the limit here is not taken over the maximal eigenvalue of $T^{(0)}(\lambda)$, because for $\Delta>1$ the maximal eigenvalue of $T^{(0)}(\lambda)$ is not in sector $n=\frac{L}{2}$. We have restricted at this point to the case of square lattices $L=N$ with even number of sites.

It is interesting to understand what mechanism gives rise to the selection rule in the case (\ref{verticalgeneral}). The product of transfer matrices in (\ref{verticalgeneral}) can be written as
\eq
\prod_{m=1}^N T^{(\theta_m)}={\left(T^{(0)}\right)}^{k_1} {\left(T^{(1)}\right)}^{k_2} \dots {\left(T^{(0)}\right)}^{k_{N-1}} {\left(T^{(1)}\right)}^{k_N} {\left(T^{(0)}\right)}^{k_{N+1}},
\label{productmatrices}
\en
where given $j$, or equivalently $\{\theta_1,\dots,\theta_N\}$, there exists a set of non negative integers $k_j$ with $\sum_{j=1}^{N+1} k_j=N$ so that (\ref{productmatrices}) is true. Hence
\begin{multline}
\prod_{m=1}^N T^{(\theta_m)}=\sum_{\langle g_j \rangle} {\left(\Lambda_{g_1}^{(0)}\right)}^{k_1} {\left(\Lambda_{g_2}^{(1)}\right)}^{k_2} \dots {\left(\Lambda_{g_{N-1}}^{(0)}\right)}^{k_{N-1}} {\left(\Lambda_{g_N}^{(1)}\right)}^{k_N} {\left(\Lambda_{g_{N+1}}^{(0)}\right)}^{k_{N+1}} \times \\
 \ket{g_1^{(0)}} \braked{g_1^{(0)}\big\vert g_2^{(1)}} \bra{g_2^{(1)}}\dots \ket{g_{N-1}^{(0)}}\braked{ g_{N-1}^{(0)} \big\vert g_{N}^{(1)}} \braked{g_{N}^{(1)} \big \vert g_{N+1}^{(0)}} \bra{g_{N+1}^{(0)}},
\end{multline}
and therefore
\bear
&&Z_{j,1}=\sum_{\langle g_j \rangle} {\left(\Lambda_{g_1}^{(0)}\right)}^{k_1} {\left(\Lambda_{g_2}^{(1)}\right)}^{k_2} \dots {\left(\Lambda_{g_{N-1}}^{(0)}\right)}^{k_{N-1}} {\left(\Lambda_{g_N}^{(1)}\right)}^{k_N} {\left(\Lambda_{g_{N+1}}^{(0)}\right)}^{k_{N+1}}  \label{zj1}\\
 &\times& \braked{g_1^{(0)}\big\vert g_2^{(1)}} \bra{g_2^{(1)}}\dots \ket{g_{N-1}^{(0)}}\braked{ g_{N-1}^{(0)} \big\vert g_{N}^{(1)}} \braked{g_{N}^{(1)} \big \vert g_{N+1}^{(0)}} \braked{g_{N+1}^{(0)} \big \vert g_1^{(0)}}. \nonumber
\ear
For $\Theta=1$, which does not satisfy the selection rule, we may take $k_1=k-1$, $k_2=1$ and $k_3=N-k$. The summation over all eigenvectors of $T^{(1)}(\lambda)$ can be written by grouping together the contribution of pairs $\Lambda_{g_2}(\lambda)$ and $-\Lambda_{g_2}(\lambda)$. This gives us
\eq
Z_{j,1}=\sum_{\langle g_1, g_2' \rangle} {\left[\Lambda_{g_1}^{(0)}\right]}^{N-1} \Lambda_{g_2}^{(1)} \bra{g_{1}^{(0)}} \Bigg[\ket{g_2^{(1)}} \bra{g_2^{(1)}}-\Pi^z \ket{g_2^{(1)}} \bra{g_2^{(1)}} \Pi^z \Bigg]\ket{g_1^{(0)}},
\en
choosing the eigenvectors of $T^{(0)}(\lambda)$ to have definite quantum number of parity, we see that $Z_{j,1}$ vanishes.

In order to study the free-energy in the thermodynamic limit, we choose the states $\ket{g_j^{(0)}}$ to have definite projection of total spin-$z$ and also $\ket{g_j^{(1)}}$ to have definite quantum number of inversion operator. We also restrict ourselves to the case of $N$ even and $L$ odd for simplicity, however the final results for different parity of $N$ and $L$ are similar to the one discussed below.

Generally, the partition function (\ref{zj1}) can be written as
\bear
&&Z_{j,1}=\sum_{\langle g_{\text{odd}}, g_{\text{even}}'\rangle} {\left(\Lambda_{g_1}^{(0)}\right)}^{k_1} {\left(\Lambda_{g_2}^{(1)}\right)}^{k_2} \dots {\left(\Lambda_{g_{N-1}}^{(0)}\right)}^{k_{N-1}} {\left(\Lambda_{g_{N}}^{(1)}\right)}^{k_{N}} {\left(\Lambda_{g_{N+1}}^{(0)}\right)}^{k_{N+1}} \nonumber\\
 &\times& \braked{g_1^{(0)}\big\vert g_2^{(1)}} \bra{g_2^{(1)}}\dots \ket{g_{N-1}^{(0)}}\braked{ g_{N-1}^{(0)} \big\vert g_{N}^{(1)}} \braked{g_{N}^{(1)} \big \vert g_{N+1}^{(0)}} \braked{g_{N+1}^{(0)} \big \vert g_1^{(0)}} \label{zj1-2}\\
 &\times& \left[1+{(-1)}^{k_2+\alpha_{g_1}+\alpha_{g_3}}\right]  \left[1+{(-1)}^{k_4+\alpha_{g_3}+\alpha_{g_5}}\right]  \dots  \left[1+{(-1)}^{k_{N}+\alpha_{g_{N-1}}+\alpha_{g_{N+1}}}\right], \nonumber
\ear
where the numbers $\alpha_j=0,1$ depending on the parity of the eigenstate $\ket{g_j^{(0)}}$. 

For the case $\Theta=2$, we have two distinct cases. The one we called consecutive inversions, is the case where there is a consecutive product of two $T^{(1)}(\lambda)$ matrices in between the product of $T^{(0)}(\lambda)$ matrices, which implies $k_1=m-2,~k_2=2,~k_3=N-m$. In the other case, we could have separated inversions, where the two $T^{(1)}(\lambda)$ matrices representing the inversions are apart. This means that $k_1=m_1-1,~~k_2=1,~~k_3=m_2-m_1-1,~~k_4=1,~~k_5=N-m_2$, $m_2-m_1 \geq 2$, and the remaining $k_j$ are all zero.

In order to illustrate, we discuss the case of consecutive inversions and we leave the case of separate inversions for the Appendix C.

The consecutive inversions in (\ref{zj1-2}) can be shortly written as,
\eq
Z_{j,1}= \sum_{g_1,g_2'} 2 {\left(\Lambda_{g_1}^{(0)}\right)}^{N-2} {\left(\Lambda_{g_2}^{(1)}\right)}^{2} {\left\vert \braked{g_1^{(0)}\big \vert g_2^{(1)}}\right\vert}^2,
\en
where the prime means summation over half of the states chosen between every pair $\ket{g^{(1)}}$ and $\Pi^z \ket{g^{(1)}}$. This case is similar to $\Theta=1$ but for a sign change due to $k_2=2$ instead of $k_2=1$.

Provided we have for large $L$ 
\bear
\left\vert \braked{g_{\max,n=L/2}^{(0)} \big \vert g_{\max}^{(1)}}\right\vert  >  \exp(-\delta L) \qquad \forall \delta>0~~~~\Delta \leq 1, \nonumber\\
\sum_{n=0}^L\left\vert \braked{g_{\max,n}^{(0)} \big \vert g_{\max}^{(1)}}\right\vert >  \exp(-\delta L) \qquad \forall \delta>0~~~~\Delta > 1,
\label{hip1}
\ear
as verified for finite lattices. This implies that
\eq
F_{j,1}=\lim_{L,N \rightarrow \infty}-\frac{1}{\beta L N} \ln \left(2 \sum_{n=0}^L {\left(\Lambda_{\max,n}^{(0)}\right)}^{N-2} {\left(\Lambda_{\max}^{(1)}\right)}^{2} {\left\vert \braked{g_{\max,n}^{(0)} \big \vert g_{\max}^{(1)}}\right\vert}^2\right)=F_{PP},
\label{fj1fpp}
\en
where the condition (\ref{hip1}) assures that the projection factor $\braked{g_{\max,n}^{(0)} \big \vert g_{\max}^{(1)}}$ does not affect free-energy in the thermodynamic limit.

As a final example of the first column $Z_{j,1}$, we would like to address the case where we have maximal alternation of $T^{(0)}(\lambda)$ and $T^{(1)}(\lambda)$. In this case we have $k_1=k_2=\dots k_n=1$ and $k_{N+1}=0$. This is equivalent to have $j=j_0=\frac{2}{3}(2^N-1)$ for even $N$. Therefore, we have
\eq
Z_{j_0,1}=\sum_{g_1,g_2',\dots,g_{N-1},g_{N}'}\prod_{m=1}^{\frac{N}{2}} \Lambda_{g_{2m-1}}^{(0)} \Lambda_{g_{2m}}^{(1)} \braked{g_{2m-1}^{(0)}\big \vert g_{2m}^{(1)}} \braked{g_{2m}^{(1)}\big \vert g_{2m+1}^{(0)}} \left(1+{(-1)}^{1+\alpha_{g_{2m-1}}+\alpha_{g_{2m+1}}}\right),
\en
where $\alpha_{g_{N+1}}=\alpha_{g_1}$. The non-zero contributions are those where
$\alpha_{g_1}=1-\alpha_{g_3}=\alpha_{g_5}=\dots=1-\alpha_{g_{N-1}}$. This is only possible when $\Theta=\frac{N}{2}$ is even, otherwise every term in $Z_{j_0,1}$ is zero. Therefore, we have that
\eq
F_{j_0,1}=\lim_{L,N \rightarrow \infty}-\frac{1}{\beta N L} \ln \sum_{n=0}^L {\left(2 \left(\Lambda^{(0)}_{\max,n} \Lambda_{\max}^{(1)}\right) {\left\vert \braked{g_{\max, n}^{(0)} \vert g_{\max}^{(1)}} \right\vert}^2 \right)}^{\frac{N}{2}}.
\en
Now the number of times that the projection $\braked{g_{\max, n}^{(0)} \vert g_{\max}^{(1)}}$ appears is comparable to $N$. The same assumption  (\ref{hip1}) tell us that the corrections are negligible and that the above free-energy equals the periodic case $F_{j_0,1}=F_{PP}$.

\subsubsection{General term}

By collecting all the results above, we see that the partition function is the product of the eigenvalues $\Lambda^{(0)}(\lambda)$ and $\Lambda^{(1)}(\lambda)$  and the projection of their eigenvectors $ \braked{g_j^{(0)}|g_k^{(1)}}$ raised to certain powers. As the leading term eigenvalues agree in the thermodynamic limit and the projection of one eigenvectors  onto another does not contribute in the thermodynamic limit, we assert that $F_{j,k}$ for all values of $\Delta$ is given in the thermodynamic limit by
\eq
F_{j,k}= -\frac{1}{\beta}\lim_{L \rightarrow \infty}\frac{\ln\left(\Lambda_{\max,n=\frac{L}{2}}^{(0)}(\lambda)\right)}{L}=F_{PP},
\en
whenever $j,k$ satisfies the selection rule. 

Finally, this implies that the free-energy of the all mixed boundary condition is equal to the free-energy for periodic boundary condition. In this context, the chance for a dependence of the entropy of the six-vertex model on the boundary conditions is the case of fixed boundary conditions, which we are going to discuss in  the next section.

\newpage

\section{The six-vertex model with fixed boundary conditions}\label{fixed}

The situation changes drastically when we consider fixed boundary conditions. In this case, the boundary choices was proven to produce different results for the free-energy and entropy in the thermodynamic limit. For convenience, we consider the case of square lattices $L=N$ throughout this section.

\subsection{Ferroelectric boundary condition}

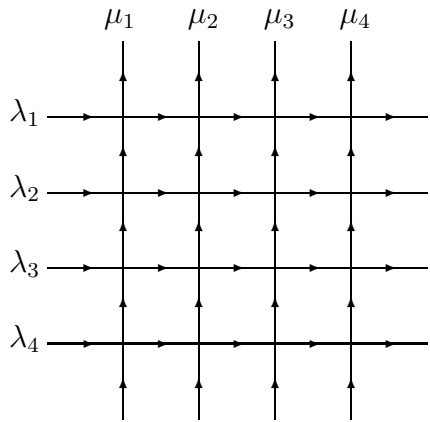
\begin{figure}[h]
\unitlength=0.5mm
\begin{center}
\begin{picture}(100,100)(-30,-10)

\multiput(-20,0)(0,20){4}{\line(1,0){100}}
\multiput(0,-20)(20,0){4}{\line(0,1){100}}
% arrow
\multiput(-8.5,0)(0,20){4}{\vector(1,0){1}}
\multiput(11.,0)(0,20){4}{\vector(1,0){1}}
\multiput(31.,0)(0,20){4}{\vector(1,0){1}}
\multiput(51.,0)(0,20){4}{\vector(1,0){1}}
\multiput(71.,0)(0,20){4}{\vector(1,0){1}}

\multiput(0,-10.)(20,0){4}{\vector(0,1){1}}
\multiput(0,11.5)(20,0){4}{\vector(0,1){1}}
\multiput(0,31.5)(20,0){4}{\vector(0,1){1}}
\multiput(0,51.5)(20,0){4}{\vector(0,1){1}}
\multiput(0,71.5)(20,0){4}{\vector(0,1){1}}

\put(-30,-1){$\lambda_4$}
\put(-30,19){$\lambda_3$}
\put(-30,39){$\lambda_2$}
\put(-30,59){$\lambda_1$}

\put(-5,85){$\mu_1$}
\put(17,85){$\mu_2$}
\put(37,85){$\mu_3$}
\put(57,85){$\mu_4$}

\end{picture}
\end{center}
\caption{The partition function $Z_N^{FE}$ for $N=4$ of the six-vertex model with ferroelectric boundary condition (FE).}
\label{fig-FE}
\end{figure}

The simplest case is the ferroelectric (FE) boundary condition (see figure \ref{fig-FE}). This boundary condition results in only one allowed physical state for any finite system size,
\eq
Z_{FE}=1,
\en
for any values of the physical parameters. This is a direct consequence of the ice rule and it was already noted in\cite{WU}.

In fact, there are four of such boundaries associated to the four possible vertices $\omega_i$ for $i=1,\dots,4$. One can obtain one from another by means of rotations of the lattice. However, for all the four case, one has that the entropy is zero ($S_{FE}=0$).

\subsection{Domain wall boundary condition}

The first non-trivial instance appeared in the context of scalar products of the Bethe states. This is the case of the partition function the with domain wall (DWBC) boundary condition \cite{KOREPIN1982},
\eq
Z_N^{DWBC}(\{\lambda\},\{\mu\})=\bra{\Downarrow}B(\lambda_N)\cdots B(\lambda_2) B(\lambda_1)\ket{\Uparrow}.
\en

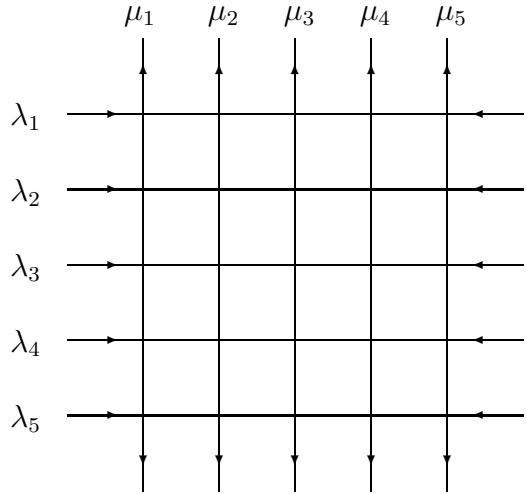
\begin{figure}[h]
\unitlength=0.5mm
\begin{center}
\begin{picture}(130,130)(-30,-10)

\multiput(-20,0)(0,20){5}{\line(1,0){120}}
\multiput(0,-20)(20,0){5}{\line(0,1){120}}
% arrow
\multiput(-7.5,0)(0,20){5}{\vector(1,0){1}}
\multiput(87.5,0)(0,20){5}{\vector(-1,0){1}}
\multiput(0,-12.5)(20,0){5}{\vector(0,-1){1}}
\multiput(0,92.5)(20,0){5}{\vector(0,1){1}}

\put(-35,-3){$\lambda_5$}
\put(-35,17){$\lambda_4$}
\put(-35,37){$\lambda_3$}
\put(-35,57){$\lambda_2$}
\put(-35,77){$\lambda_1$}

\put(-5,105){$\mu_1$}
\put(17,105){$\mu_2$}
\put(37,105){$\mu_3$}
\put(57,105){$\mu_4$}
\put(77,105){$\mu_5$}

\end{picture}
\end{center}
\caption{The partition function $Z_N^{DWBC}$ for $N=5$ of the six-vertex model with domain wall boundary condition.}
\label{fig-DWBC}
\end{figure}

The above partition function can be written in a determinant form \cite{KOREPIN1992}, which was useful to the understanding of the thermodynamic limit of the six-vertex model with DWBC. The results for the free-energy and entropy were surprisingly different in comparison with the periodic boundary\cite{KOREPIN2000,ZINNJUSTIN}. These results and its finite size corrections were rigorously proven \cite{BLEHER}.

The partition function $Z_N^{DWBC}$ is one of the fixed boundary conditions that one has inside the case of boundary $Z_{AA}$. Actually, there are two equivalent domain wall partition function inside the double anti-periodic boundary case, which are trivially related by rotation of the lattice.

The entropy at infinite temperature $S_{DWBC}=\frac{1}{2} \ln\left(\frac{3^3}{2^4}\right)\approx 0.261624$ is smaller than the periodic case.

One could rise the question if are there other boundary conditions which result in an entropy different from periodic case.

\subsection{DWBC descendent}

\begin{figure}[h]
\unitlength=0.5mm
\begin{center}
\begin{picture}(130,130)(-30,-10)

\multiput(-20,0)(0,20){5}{\line(1,0){120}}
\multiput(0,-20)(20,0){5}{\line(0,1){120}}
% arrow
\multiput(-7.5,20)(0,20){3}{\vector(1,0){1}}
\multiput(87.5,20)(0,20){3}{\vector(-1,0){1}}
\multiput(20,-12.5)(20,0){3}{\vector(0,-1){1}}
\multiput(20,92.5)(20,0){3}{\vector(0,1){1}}

\put(-17,-5){$s_3$}
\put(-7,-16){$\bar{s}_3$}
\put(-17,82){$s_1$}
\put(-8,90){$s_1$}

\put(90,-6){$s_4$}
\put(81,-16){$s_4$}
\put(90,82){$s_2$}
\put(81,90){$\bar{s}_2$}

\put(-35,-3){$\lambda_5$}
\put(-35,17){$\lambda_4$}
\put(-35,37){$\lambda_3$}
\put(-35,57){$\lambda_2$}
\put(-35,77){$\lambda_1$}

\put(-5,105){$\mu_1$}
\put(17,105){$\mu_2$}
\put(37,105){$\mu_3$}
\put(57,105){$\mu_4$}
\put(77,105){$\mu_5$}

\end{picture}
\end{center}
\caption{The partition function $Z_N^{dDWBC}$ of the six-vertex model with descendent domain wall boundary condition, where $s_i=\uparrow,\downarrow$ or $\rightarrow,\leftarrow$ and $\bar{s}_i$ is its reverse.}
\label{fig-dDWBC}
\end{figure}
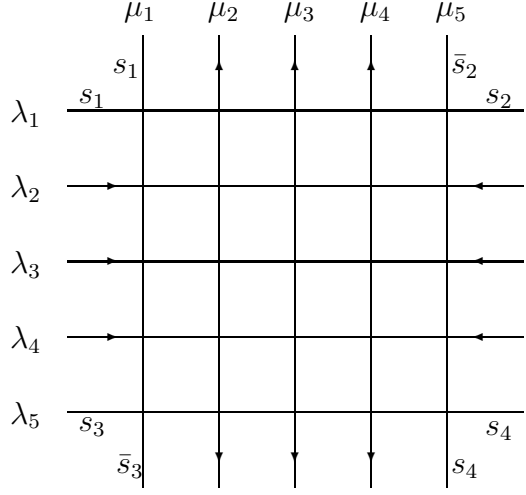

In order to address to the question about the existence of other special fixed boundaries, we extensively investigated the number of configurations of other boundary conditions. Although the complete classification of the boundary in groups of similar pattern has eluded so far, we have found some interesting cases. Surprisingly, we found that there is a family of boundary condition which share exactly the same number of configuration as the domain wall boundary. This equivalence of the number of configurations occurs at infinite temperature. We call these boundary as descendent of the domain wall boundary condition (dDWBC). In Figure \ref{fig-dDWBC}, we give an example for $N=5$.

The dDWBC that we have been able to classify and know to persist for larger $N$ can be treated by integrability tools.
We use the algebraic Bethe ansatz to derive a recurrence relation for the $Z_N^{dDWBC}$, which establish a relation with the conventional domain wall partition function $Z_{N-1}^{DWBC}$ of a smaller lattice. This is done along the same lines as \cite{PRONKO}, but here we obtain a relation between different partition functions $Z_N^{dDWBC}$ and $Z_{N-1}^{DWBC}$. The partition function $Z_N^{dDWBC}$ (see Figure \ref{fig-dDWBC}) can be written as,
\bear
&&Z_N^{dDWBC}(\{\lambda\},\{\mu\})= \label{PF-dDWBC}\\
&=&\bra{\bar{s}_3\Downarrow_{N-2}s_4}{\left({\cal T}_{\cal A}(\lambda_N)\right)}_{s_3,s_4} B(\lambda_{N-1})\cdots B(\lambda_2) {\left({\cal T}_{\cal A}(\lambda_1)\right)}_{s_1,s_2} \ket{s_1\Uparrow_{N-2} \bar{s}_2}, \nonumber
\ear
where $s_i=\uparrow, \downarrow$ or $\rightarrow,\leftarrow$ and $\bar{s}_i$ is its reverse.

In order to illustrate this, we shall take $s_1=s_2=s_4=\downarrow$ , and $s_3= \uparrow$ as in
\eq
Z_N(\{\lambda\},\{\mu\})= \bra{\Downarrow_N} B(\lambda_N) \dots B(\lambda_{2}) D(\lambda_1) \ket{\downarrow  \Uparrow_{N-1}}.
\en
Using the two-site model decomposition, where the monodromy matrix is decomposed into two parts,
\eq
{\cal T}_{\cal A}(\lambda)=\left(\begin{array}{cc}
                            A_{N-1}(\lambda) & B_{N-1}(\lambda) \\
                            C_{N-1}(\lambda) & D_{N-1}(\lambda)
                           \end{array}\right) \left(\begin{array}{cc}
                            A_1(\lambda) & B_1(\lambda) \\
                            C_1(\lambda) & D_1(\lambda)
                           \end{array}\right),
\en
we can derive the following relation,
\begin{align}
Z_N(\{\lambda\},\{\mu\})=\sum_{j=1}^N r_j \bra{\Downarrow_{N-1}} \prod_{k=j+1}^{\stackrel{\curvearrowleft}{N}} B_{N-1}(\lambda_k) D_{N-1}(\lambda_j) \prod_{k=1}^{\stackrel{\curvearrowleft}{j-1}} B_{N-1}(\lambda_k) \ket{\Uparrow_{N-1}},\label{rec1}
\end{align}
where
\begin{align}
r_1&=a(\lambda_1-\mu_1) \prod_{m=2}^N b(\lambda_m-\mu_1), \qquad r_2=\frac{c(\lambda_1-\mu_1)  c(\lambda_2-\mu_1)}{a(\lambda_1-\mu_1) b(\lambda_2-\mu_1)} r_1, \nonumber \\
r_j&= \frac{a(\lambda_{j-1}-\mu_1) c(\lambda_j-\mu_1)}{c(\lambda_{j-1}-\mu_1) b(\lambda_j- \mu_1)} r_{j-1} \qquad j=3, \dots,N.
\end{align}
From now on in this section, we drop for convenience the indices $N-1$ of the operators $B(\lambda)$ and $D(\lambda)$, which acts in space $\prod_{m=2}^N V_m$.
We use the Yang-Baxter algebra (\ref{fundrel}) to move the $D(\lambda)$ operators to the right. In order to do so, we need the commutativity properties of the $B(\lambda)$ and the relation
\eq
D(\lambda_j) \prod_{m=1}^{\stackrel{\curvearrowleft}{j-1}} B(\lambda_m) = \sum_{k=1}^j \beta_{j k} \prod_{\stackrel{m=1}{m \neq k}}^{\stackrel{\curvearrowleft}{j}} B(\lambda_m) D(\lambda_k),
\en
where
\bear
\beta_{jk}=\begin{cases}
\displaystyle-\frac{c(\lambda_j-\lambda_k)}{b(\lambda_j-\lambda_k)} \prod_{\stackrel{i=1}{i \neq k}}^j \frac{a(\lambda_k-\lambda_i)}{b(\lambda_k-\lambda_i)}, & k \neq j, \\
\displaystyle\prod_{i=1}^{j-1} \frac{a(\lambda_j-\lambda_i)}{b(\lambda_j-\lambda_i)} , & k=j.
\end{cases}
\ear

As the state $\ket{\Uparrow}$ is an eigenstate of operator $D(\lambda)$, we are left with the following ``recursion'' relation,
\eq
Z_N(\{\lambda\},\{\mu\}) = \sum_{k=1}^N Z_{N-1}^{DWBC}(\{\lambda\}\setminus \lambda_k,\{\mu\} \setminus \mu_1)\left[(b(\lambda_k))^{N-1} \sum_{j=k}^N r_j \beta_{jk}\right].
\label{reccurencex}
\en

It is possible to obtain a concise expression for the partition function in terms of a determinant formula. In doing so, we replace the determinant formula for the domain wall partition function \cite{KOREPIN1992} given by,
\eq
Z_N^{DWBC} (\{\lambda\},\{\mu\})= f_N(\{\lambda\},\{\mu\}) \det{\big[\rho(\lambda_i,\mu_j)\big]}_{i=1,\dots,N}^{j=1, \dots, N},
\label{Zdwbc}
\en
in the relation (\ref{reccurencex}). We finally obtain that
\eq
Z_N(\{\lambda\},\{\mu\})= \left|
                              \begin{array}{ccccc}
                                \delta_1 & \rho(\lambda_1,\mu_2) & \rho(\lambda_1,\mu_3) & \dots & \rho(\lambda_1,\mu_N) \\
                                \delta_2 & \rho(\lambda_2,\mu_2) & \rho(\lambda_2,\mu_3) & \dots & \rho(\lambda_2,\mu_N) \\
                                \vdots   & \vdots                & \vdots                & \ddots & \vdots                \\
                                \delta_N & \rho(\lambda_N,\mu_2) & \rho(\lambda_N,\mu_3) & \dots & \rho(\lambda_N,\mu_N)   \\
                              \end{array}
                            \right|,
\en
where $\delta_k$, $\rho(\lambda,\mu)$ and $f_N(\{\lambda\},\{\mu\})$ are given by
\bear
\delta_k &=& {(-1)}^{1+k} f_{N-1}(\{\lambda\}\setminus \lambda_k,\{\mu\} \setminus \mu_1) b^{N-1}(\lambda_k) \sum_{j=k}^N r_j \beta_{jk}, \\
\rho(\lambda,\mu)&=&\frac{c(\lambda-\mu)}{a(\lambda-\mu) b(\lambda-\mu)},
\label{rhofu}
\ear
\begin{align}
&f_N(\{\lambda\},\{\mu\})= \nonumber \\
&=\frac{\displaystyle\prod_{\stackrel{i,j=1}{i<j}}^N\frac{(c_{ij}c_{ji}b_{i i}b_{jj}+c_{i i} c_{j j} a_{i j} a_{j i}) (c_{ii} c_{jj} b_{i j} b_{j i}+c_{i j} c_{j i} a_{i i} a_{jj})}{\rho_{ii} \rho_{jj} (c_{i j} c_{j i} b_{i i} b_{jj}+c_{i i} c_{j j} a_{i j} a_{j i})-\rho_{i j} \rho_{j i}(c_{i j} c_{j i} a_{i i} a_{j j}+c_{ i i} c_{j j} b_{i j} b_{j i})}}{\prod_{i=1}^N {(a_{ii}b_{ii})}^{N-2}},  
\label{fcoef}
\end{align}
and we have denoted $a_{ij}= a(\lambda_i-\mu_j)$ and so on. Interesting enough, the representation of the partition function with domain wall boundary condition given in the expression (\ref{Zdwbc}) with (\ref{rhofu}-\ref{fcoef}) holds true for all physical regimes, which means all values of $\Delta$.

Taking the homogeneous limit and  and setting all Boltzmann weights to unity (the ice-point), we obtain as expected the number of alternating sign matrices \cite{KUPERBERG1996}. Therefore, this boundary produce the same entropy as the domain wall boundary \cite{KOREPIN2000}.

The reversal of corner arrows which defines the partition function (\ref{PF-dDWBC}) corresponds to exchange Boltzmann weight pairs $\{\omega_1,\omega_5\}$ or $\{\omega_2,\omega_6\}$ for up-left corner, $\{\omega_1,\omega_6\}$ or $\{\omega_2,\omega_5\}$ for down-right corner, $\{\omega_3,\omega_6\}$ or $\{\omega_4,\omega_5\}$ for up-right corner and $\{\omega_3,\omega_5\}$ or $\{\omega_4,\omega_6\}$ for down-left corner. Therefore, whatever configuration of compatible internal arrows is chosen, the new partition function is readily obtained by making the appropriate exchange of the Boltzmann weights of the corners.

It is worth to note that this invariance by the exchange of the corner arrows is not restricted to domain wall boundaries. We can also find similar invariance in any other fixed boundary.

\subsection{Merge of DWBC and FE boundary}

We have also discovered another class of boundary condition which are related with the domain wall boundary conditions. However, in this case the number of physical states and therefore the entropy is smaller than the domain wall boundary.

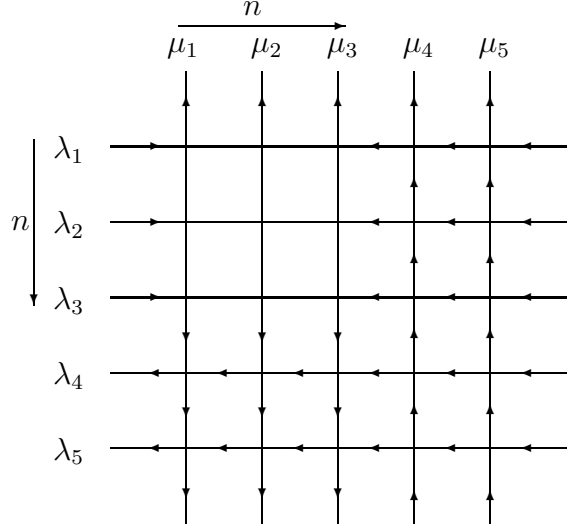
\begin{figure}[ht]
\unitlength=0.5mm
\begin{center}
\begin{picture}(130,130)(-30,-10)

\multiput(-20,0)(0,20){5}{\line(1,0){120}}
\multiput(0,-20)(20,0){5}{\line(0,1){120}}
% arrow
\multiput(-7.5,40)(0,20){3}{\vector(1,0){1}}
\multiput(89,0)(0,20){5}{\vector(-1,0){1}}
\multiput(0,-12.5)(20,0){3}{\vector(0,-1){1}}
\multiput(0,92.5)(20,0){5}{\vector(0,1){1}}

\put(-40,82){\vector(0,-1){44}}
\put(-46,58){$n$}
\put(-2,112){\vector(1,0){44}}
\put(15,115){$n$}

\multiput(-9,0)(0,20){2}{\vector(-1,0){1}}
\multiput(60,-12)(20,0){2}{\vector(0,1){1}}

\multiput(69,0)(0,20){5}{\vector(-1,0){1}}
\multiput(49,0)(0,20){5}{\vector(-1,0){1}}
\multiput(29,0)(0,20){2}{\vector(-1,0){1}}
\multiput(9,0)(0,20){2}{\vector(-1,0){1}}

\multiput(0,9)(20,0){3}{\vector(0,-1){1}}
\multiput(0,29)(20,0){3}{\vector(0,-1){1}}

\multiput(60,11)(20,0){2}{\vector(0,1){1}}
\multiput(60,31)(20,0){2}{\vector(0,1){1}}
\multiput(60,51)(20,0){2}{\vector(0,1){1}}
\multiput(60,71)(20,0){2}{\vector(0,1){1}}

\put(-35,-3){$\lambda_5$}
\put(-35,17){$\lambda_4$}
\put(-35,37){$\lambda_3$}
\put(-35,57){$\lambda_2$}
\put(-35,77){$\lambda_1$}

\put(-5,105){$\mu_1$}
\put(17,105){$\mu_2$}
\put(37,105){$\mu_3$}
\put(57,105){$\mu_4$}
\put(77,105){$\mu_5$}

\end{picture}
\end{center}
\caption{The partition function $Z_N^{fDWBC}$ for $N=5$ of the six-vertex model whose boundary are a mixture of the domain wall boundary condition and ferroelectric boundary (fDWBC).}
\label{fig-fDWBC}
\end{figure}

These boundaries are actually a mixture or a merge of domain wall and ferroelectric boundary conditions. We choose an integer number $n$ between $0$ and $N$. Starting from the upper-left corner, we fill the first $n$ boundary row and column edges with arrows in the same way we would fill the domain wall boundary of type $\omega_5$, see Figure \ref{fig-fDWBC}. The opposite edges of these are also filled with the respective arrows of the opposite edges of the domain wall boundary condition. So far, we have used the arrows configuration of a domain wall boundary with lattice size $n$ to fill our boundary of lattice size $N$. The remaining arrows are filled in the same way, but using boundary arrows of the ferroelectric boundary condition of type $\omega_4$.

Considering this boundary, we have partially frozen the arrow configurations of the lattice in a similar way as the ferroelectric boundary. The only difference is due to the $n\times n$ sublattice at the upper-left corner. This implies we are left with a domain wall partition function of size $n$, which means $Z_N^{fDWBC}= \prod_{i j \not \in ~n \times n} b(\lambda_i-\mu_j)\times Z_n^{DWBC}$. Therefore, we see that the entropy at infinity temperature is given by
\eq
S_{fDWBC}=\lim_{N \rightarrow \infty }{\left(\frac{n}{N} \right)}^2 S_{DWBC}.
\en
For a suitably chosen sequence $n(N)$, one can obtain any value of entropy $S$, such that $S_{FE} \leq S \leq S_{DWBC}$.

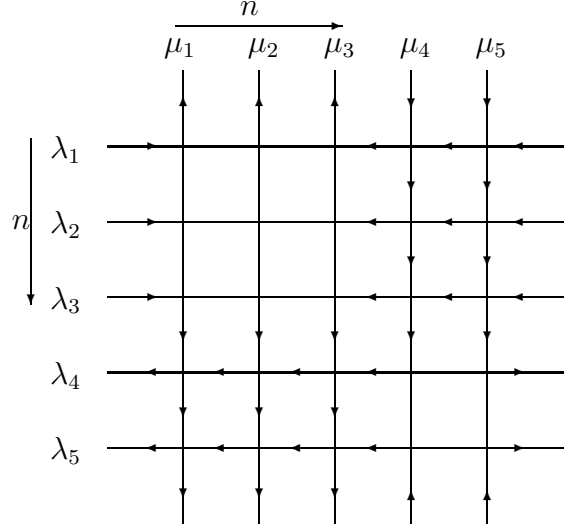
\begin{figure}[ht]
\unitlength=0.5mm
\begin{center}
\begin{picture}(130,130)(-30,-10)

\multiput(-20,0)(0,20){5}{\line(1,0){120}}
\multiput(0,-20)(20,0){5}{\line(0,1){120}}
% arrow

\multiput(-7.5,40)(0,20){3}{\vector(1,0){1}}
\multiput(-9,0)(0,20){2}{\vector(-1,0){1}}

\multiput(87.5,40)(0,20){3}{\vector(-1,0){1}}
\multiput(89,0)(0,20){2}{\vector(1,0){1}}

\multiput(0,-12.5)(20,0){3}{\vector(0,-1){1}}
\multiput(60,-12)(20,0){2}{\vector(0,1){1}}

\multiput(0,92.5)(20,0){3}{\vector(0,1){1}}
\multiput(60,91.0)(20,0){2}{\vector(0,-1){1}}

\put(-40,82){\vector(0,-1){44}}
\put(-45,58){$n$}
\put(-2,112){\vector(1,0){44}}
\put(15,115){$n$}

\multiput(69,40)(0,20){3}{\vector(-1,0){1}}
\multiput(49,40)(0,20){3}{\vector(-1,0){1}}
\multiput(29,0)(0,20){2}{\vector(-1,0){1}}
\multiput(9,0)(0,20){2}{\vector(-1,0){1}}
\multiput(49,0)(0,20){2}{\vector(-1,0){1}}

\multiput(0,9)(20,0){3}{\vector(0,-1){1}}
\multiput(0,29)(20,0){3}{\vector(0,-1){1}}

\multiput(60,29)(20,0){2}{\vector(0,-1){1}}
\multiput(60,49)(20,0){2}{\vector(0,-1){1}}
\multiput(60,69)(20,0){2}{\vector(0,-1){1}}

\put(-35,-3){$\lambda_5$}
\put(-35,17){$\lambda_4$}
\put(-35,37){$\lambda_3$}
\put(-35,57){$\lambda_2$}
\put(-35,77){$\lambda_1$}

\put(-5,105){$\mu_1$}
\put(17,105){$\mu_2$}
\put(37,105){$\mu_3$}
\put(57,105){$\mu_4$}
\put(77,105){$\mu_5$}

\end{picture}
\end{center}
\caption{The partition function $Z_N^{f2DWBC}$ for $N=5$ of the six-vertex model whose boundary are a merge of two domain wall boundary condition and ferroelectric boundary (f2DWBC).}
\label{fig-fDWBC2}
\end{figure}

Similarly, we can construct another kind of boundary intimately related to domain wall boundary condition. This is obtained by filling the first $n$ arrows using domain wall boundary condition of type $\omega_5$ as described before. Additionally, we fill the remaining arrows with the domain wall boundary condition of type $\omega_6$, as described in the Figure \ref{fig-fDWBC2}. This boundary condition obtained by the merge of two domain wall boundary has its counterpart in the context of the generalization of alternating sign matrices \cite{BRUALDI}.

As a result we obtain a lattice with frozen configurations, except for two square sublattices of sizes $n$ and $N-n$. In this case, the partition function $Z_N^{f2DWBC}$ is a product of two domain wall partition functions of size $n$ and $N-n$ and the weights $a(\lambda_i- \mu_j)$ with $(i,j)$ outside the two domain wall square sublattices, 
\eq
\frac{Z_N^{f2DWBC}}{\prod_{\substack{i=1,\dots,n\\ j=n+1,\dots N}} a(\lambda_i- \mu_j) a(\lambda_j-\mu_i)}= Z_n^{DWBC}(\{\lambda\}_{1}^{n},\{\mu\}_{1}^n) Z_{N-n}^{DWBC}(\{\lambda\}_{n+1}^{N},\{\mu\}_{n+1}^{N}).
\en
Therefore, we see that the entropy at infinite temperature is given by
\eq
S_{f2DWBC}=\lim_{N \rightarrow \infty} \left(1-2 \frac{n}{N} \left(1-\frac{n}{N}\right)\right)S_{DWBC},
\en
whose minimum value is given by the sequence $n(N)=\lceil\frac{N}{2}\rceil$
\eq
S_{f2DWBC}^{(\min)}=\frac{1}{2}S_{DWBC},
\en
which is the case where we have the merge of two domain wall boundary condition and the ferroelectric boundary. This again produces entropies smaller than the usual domain wall boundary condition. 

\subsection{N\'eel boundary condition}

\begin{figure}[h]
\unitlength=0.5mm
\begin{center}
\begin{picture}(100,100)(-30,-10)

\multiput(-20,0)(0,20){4}{\line(1,0){100}}
\multiput(0,-20)(20,0){4}{\line(0,1){100}}
% arrow
\multiput(-10.5,0)(0,40){2}{\vector(-1,0){1}}
\multiput(-8.5,20)(0,40){2}{\vector(1,0){1}}
\multiput(69.,0)(0,40){2}{\vector(-1,0){1}}
\multiput(71.,20)(0,40){2}{\vector(1,0){1}}

\multiput(20,-12.)(40,0){2}{\vector(0,-1){1}}
\multiput(0,-10.)(40,0){2}{\vector(0,1){1}}

\multiput(20,69.5)(40,0){2}{\vector(0,-1){1}}
\multiput(0,71.5)(40,0){2}{\vector(0,1){1}}

\put(-30,-1){$\lambda_4$}
\put(-30,19){$\lambda_3$}
\put(-30,39){$\lambda_2$}
\put(-30,59){$\lambda_1$}

\put(-5,85){$\mu_1$}
\put(17,85){$\mu_2$}
\put(37,85){$\mu_3$}
\put(57,85){$\mu_4$}

\end{picture}
\end{center}
\caption{The partition function $Z_N^{NE}$ for $N=4$ of the six-vertex model with N\'eel boundary condition (NE).}
\label{fig-NE}
\end{figure}
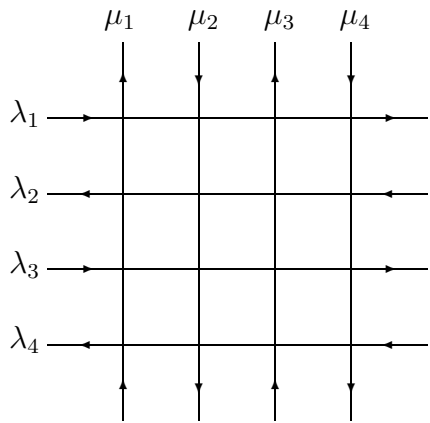

The last case we are considering is what we called N\'eel boundary condition or anti-ferroelectric boundary. This is the case where we have the alternation of the arrows along the boundaries, see Figure \ref{fig-NE}. We can find this pattern of fixed boundary inside the case periodic boundary condition along both direction $Z_{PP}$ for even lattices. There is an analogue state for odd $N$, which is one of the states of the anti-periodic boundary case $Z_{AA}$. 

In contrast with the case of ferroelectric boundary condition which allows for only one possible state, the N\'eel boundary is the one which allows for the largest number of configurations. This is due to the ice rule, which restricts the arrow reversal along rows and columns through the vertices $\omega_5$ and $\omega_6$. The reversal of arrows along the lines implies in an arrow reversal along the column and vice-versa. As a consequence of the arrow alternation in the boundary, the N\'eel boundary allows the larger number of arrow reversals along the boundary and this propagates to the bulk.

We computed the number of states of all possible boundaries for lattices up to $N=6$ and the N\'eel boundary has the largest number of physical states among all fixed boundary conditions. Naturally, one has that the number of configuration of the N\'eel boundary $\Omega_{NE}$ is smaller than the periodic boundary case $\Omega_{PP}$. Nevertheless, we can also estimate an upper bound value for $\Omega_{PP}$ assuming that all possible combination $2^{2N}$ of arrows on the boundary produce the same number of configuration $\Omega_{NE}$. This provides the following relation,
\eq
\Omega_{NE} \leq \Omega_{PP} \leq 2^{2 N} \Omega_{NE},
\label{NEPP}
\en
which implies that their entropies coincide $S_{NE}=S_{PBC}$ at infinite temperature.

Alternatively, one can express the partition function in terms of the monodromy matrix elements, which holds for all temperature,
\eq
Z_N^{NE}(\{\lambda\},\{\mu\})=\bra{\uparrow \downarrow \dots \uparrow \downarrow}D(\lambda_N) A(\lambda_{N-1})\cdots  D(\lambda_{2}) A(\lambda_1) \ket{\uparrow \downarrow \dots \uparrow \downarrow},
\en
however, it should be noted that the operators $A(\lambda)$, $D(\lambda)$ and $D(\lambda) A(\lambda)$ are not orthogonal, except in the ice-point where the product $D(\lambda) A(\lambda)$ becomes orthogonal. Furthermore, the partition function $Z_N^{NE}$ is not a symmetric function of the spectral parameters due to the fact that these operators do not commute. Therefore, we do not have a simple determinant formula for the partition function. Even though the $D(\lambda) A(\lambda)$ still commutes with total spin-$z$ operator, we also could not obtain its eigenvalues by analytical means.

Nevertheless, we can compute the number of configurations for the N\'eel boundary for finite lattices. We have done that for lattices up to $N=20$. The number of configuration are given in the Table \ref{tab-numbNE}.

\begin{table}
\begin{center}
\begin{tabular}{|l|l|}
\hline
$N$ & number of states \\
\hline
\footnotesize 1 & \footnotesize 1 \\
\footnotesize 2 & \footnotesize 2 \\
\footnotesize 3 & \footnotesize 7 \\
\footnotesize 4 & \footnotesize 64 \\
\footnotesize 5 & \footnotesize 1322 \\
\footnotesize 6 & \footnotesize 64914 \\
\footnotesize 7 & \footnotesize 7474305 \\
\footnotesize 8 & \footnotesize 2033739170 \\
\footnotesize 9 & \footnotesize 1305583070738 \\
\footnotesize 10 & \footnotesize 1981880443295788 \\
\footnotesize 11 & \footnotesize 7111657020627320662 \\
\footnotesize 12 & \footnotesize 60382974032926242142168 \\
\footnotesize 13 & \footnotesize 1213039653244899907872180826 \\
\footnotesize 14 & \footnotesize 57687270950680153355854587442676 \\
\footnotesize 15 & \footnotesize 6494209210696211480439308528411663853 \\
\footnotesize 16 & \footnotesize 1731204438495421321106461120147832169010790 \\
\footnotesize 17 & \footnotesize 1092829001103470428650265862752651675963745966742 \\
\footnotesize 18 & \footnotesize 1633892840599915791908254127642749411000513938128114064 \\
\footnotesize 19 & \footnotesize 5785898354977820698935460290451680551971080689572072829375890 \\
\footnotesize 20 & \footnotesize 48534629904275880189653389798729712740901732087151544103619504415896 \\
\hline
\end{tabular}
\caption{Number of configurations for N\'{e}el boundary condition.}
\label{tab-numbNE}
\end{center}
\end{table}

\begin{figure}
\begin{center}
\includegraphics[width=0.8\columnwidth]{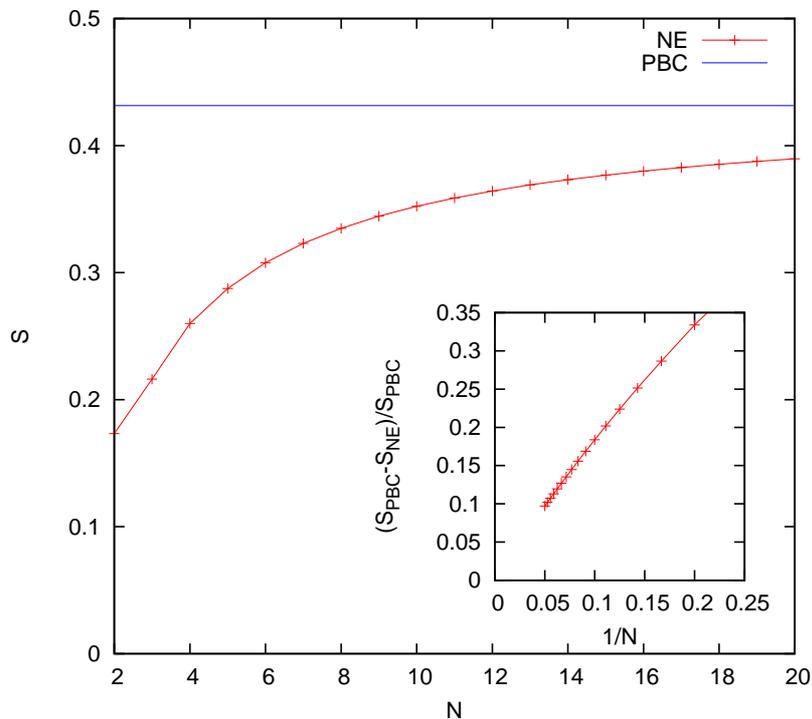}
\end{center}
\caption{(Color online) Entropy as a function of the lattice size $N$ for the N\'eel boundary in comparison with the entropy for periodic boundary condition in the thermodynamic limit (\ref{entrPBC}).}
\label{figurex}
\end{figure}

While the case of domain wall boundary is well known to have a product formula for the number of states involving factorials, it seems that there is no such formula for the case of N\'eel boundary condition. Therefore, we are not able to compute the entropy exactly due to the fact the this boundary does not seem to be integrable. However, one can compare the N\'eel and periodic boundary condition entropies and extract the large $N$ behavior (see Figure \ref{figurex}). The entropy seems to behave as $S_{NE}=S_{PBC} (1 -\frac{\gamma}{N})$, where $\gamma \sim 2$, which confirms the reasoning in (\ref{NEPP}).

The first four numbers of states for the N\'eel boundary condition given in the Table \ref{tab-numbNE} has also recently appeared in the context of generalized alternating sign matrices\cite{BRUALDI}. In fact, there is a correspondence between number of physical configurations of the six-vertex model and the generalized alternating sign matrices. The N\'{e}el boundary condition ($N$ even) address the problem of counting the number of matrices of the following type: each nonzero entry can be either $+1$ or $-1$; the odd rows or columns, if they have any nonzero entry, then the nonzero entries  must start with $+1$ and end with $-1$ as we fill the row(column) in crescent direction of indices; the even rows or columns, if they have any nonzero entry, then the nonzero entries must start with $-1$ and end with $+1$. This is to be contrasted with the case of alternating sign matrices, where the nonzero entries should always start with $+1$ and also end with $+1$ \cite{ASM}.

It was shown in \cite{BRUALDI} the sufficient and necessary conditions for existence of generalized alternating sign matrices of specific types. In the context of vertex models, this should correspond to the determination of the non-trivial fixed boundary conditions of the six-vertex model. 
The fact that the number of states for lattices up to $N=3$ coincides with the number of alternating sign matrices raised the possibility for the agreement  between the number of states of the DWBC and the N\'eel boundary, which was soon dismissed by looking at $N=4$ \cite{BRUALDI}.
Whereas in the context of vertex models discussed here, it is clear that the DWBC has fewer configurations than the N\'eel boundary condition.

\section{Merge of DWBC and free or N\'eel boundaries}

\begin{figure}[h]
\unitlength=0.5mm
\begin{center}
\begin{picture}(130,130)(-30,-10)

\multiput(-15,0)(0,15){7}{\line(1,0){120}}
\multiput(0,-15)(15,0){7}{\line(0,1){120}}
% arrow
\multiput(-7.5,0)(0,15){2}{\vector(1,0){1}}
\multiput(-7.5,75)(0,15){2}{\vector(1,0){1}}

\multiput(97.5,0)(0,15){2}{\vector(-1,0){1}}
\multiput(97.5,75)(0,15){2}{\vector(-1,0){1}}

\multiput(0,-10)(15,0){2}{\vector(0,-1){1}}
\multiput(75,-10)(15,0){2}{\vector(0,-1){1}}
\multiput(0,97.5)(15,0){2}{\vector(0,1){1}}
\multiput(75,97.5)(15,0){2}{\vector(0,1){1}}

\put(-35,-3){$\lambda_7$}
\put(-35,12){$\lambda_6$}
\put(-35,27){$\lambda_5$}
\put(-35,42){$\lambda_4$}
\put(-35,57){$\lambda_3$}
\put(-35,72){$\lambda_2$}
\put(-35,87){$\lambda_1$}

\put(-5,115){$\mu_1$}
\put(12,115){$\mu_2$}
\put(27,115){$\mu_3$}
\put(42,115){$\mu_4$}
\put(57,115){$\mu_5$}
\put(72,115){$\mu_6$}
\put(87,115){$\mu_7$}

\put(-95,42){$\displaystyle\sum_{s_i,\hat{s}_i=\uparrow,\downarrow}\sum_{r_i,\hat{r}_i=\rightarrow,\leftarrow}$}

\put(-10,25){$r_5$}
\put(-10,40){$r_4$}
\put(-10,55){$r_3$}

\put(95,23){$\hat{r}_5$}
\put(95,38){$\hat{r}_4$}
\put(95,53){$\hat{r}_3$}

\put(20,95){$s_3$}
\put(36,95){$s_4$}
\put(52,95){$s_5$}

\put(20,-10){$\hat{s}_3$}
\put(36,-10){$\hat{s}_4$}
\put(52,-10){$\hat{s}_5$}

\end{picture}
\end{center}
\caption{The partition function $Z_N^{DW-free}$ for $N=7$ of the six-vertex model with the mixture of domain wall boundary conditions and free boundary conditions along $n=3$ central edges, where $s_i,\hat{s}_i=\uparrow,\downarrow$ and $r_i,\hat{r}_i=\rightarrow,\leftarrow$.}
\label{fig-FREE}
\end{figure}
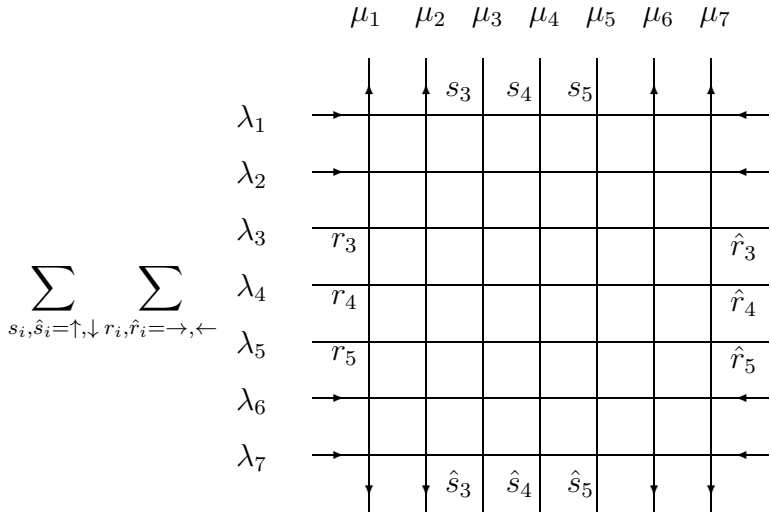

In order to address to cases where the entropy varies from $S_{DWBC}$ to $S_{PBC}$, we propose the merge of the domain wall boundary conditions with other boundary conditions which provides a large number of additional configurations. This is the case of N\'eel boundary or free boundary condition. Therefore, it appears natural to propose that the merge of domain wall boundary with free boundary condition, e.g. along $n$ central edges (see Figure \ref{fig-FREE}). This implies that the entropy may vary from its value for domain wall boundary $S_{DWBC}$ to the free boundary value $S_{free}=S_{PBC}$. Likewise, one could merge domain wall and N\'eel boundary, which again would produce entropies in the interval $S_{DWBC} < S  < S_{NE}=S_{PBC}$.

Although the conception of such examples is quite natural, the precise computation of the entropy for these examples has eluded us so far.

\section{The six-vertex at $\Delta=1$ with singular toroidal boundary condition}\label{6vsing}

In the isotropic point ($\Delta=1$) the symmetry increases, which implies that the ${\cal G}$ matrix could be any $2\times 2$ matrix
\eq
{\cal G}=\left(\begin{array}{cc}
			    g_{11} & g_{12}  \\
			    g_{21} & g_{g22}
                           \end{array}\right).
\en
This bigger symmetry at the isotropic point was used to study general toroidal boundary conditions \cite{RIBEIRO1,RIBEIRO2}. It was noted in \cite{RIBEIRO1} that the transfer matrix becomes defective when the boundary matrix ${\cal G}$ becomes singular. This implies that the transfer matrix has fewer than $2^N$ eigenvectors. Actually, it has $N+1$ eigenvectors \cite{RIBEIRO1},
\eq
\ket{\phi}^{(n)}=\bigotimes_{i=1}^{n}
{ -\frac{g_{22}}{g_{21}} \choose 1}_i
\bigotimes_{i=n+1}^{N}
{\frac{g_{11}}{g_{21}} \choose 1}_i
,\qquad n=0,1,\dots,N.
\en
and its eigenvalues are given by
\eq
\Lambda^{(n)}(\lambda)=(g_{11}+g_{22})[a(\lambda)]^{N-n}[b(\lambda)]^{n},
\label{eing-sing}
\en
whose degeneracy is $d_n=\frac{N!}{(N-n)! n!}$. Therefore, the effect of tuning the boundary parameter such that $\cal G$ becomes singular has dramatically changed the Hilbert space and the transfer matrix has the following Jordan decomposition
\eq
T(\lambda)= \diag(J_0,J_1,\cdots,J_N)
\label{jordan}
\en
where $J_n$ is the $d_n \times d_n$ Jordan matrix given by\cite{RIBEIRO1},
\eq
J_n=\left(\begin{array}{ccccc}
\Lambda^{(n)}(\lambda) & 1 & 0 & \cdots & 0 \\
0 & \Lambda^{(n)}(\lambda) & 1 & \cdots & 0 \\
\vdots & \vdots & \vdots & \ddots & \vdots \\
0 & 0 & 0 & \cdots & \Lambda^{(n)}(\lambda)
\end{array}\right)_{d_n \times d_n}.
\en

We could ask if the thermodynamic properties of the partition function would remain unchanged even in the case of singular boundary. In order to address to this point, we have to build up the partition with the full matrix $\cal G$. This is obtained by the successive multiplication of the transfer matrix $T(\lambda)=\tr_{\cal A}{[{\cal G}_{\cal A} {\cal T}_{\cal A}(\lambda)]}$,
\bear
Z&=&\tr_{V}\left[(T(\lambda))^N\right]=\sum_{n=0}^{N}d_n \left(\Lambda^{(n)}(\lambda)\right)^N, \\
&=&(g_{11}+g_{22})^N [a(\lambda)^N + b(\lambda)^N]^N,
\ear
where we have explicitly used the expressions (\ref{eing-sing}-\ref{jordan}). Taking the thermodynamic limit $F=-\frac{1}{\beta}\lim_{N,N\rightarrow \infty}\frac{1}{N N}\ln{Z}$, we obtain that the free-energy
\eq
e^{-\beta F}=\max{(a,b)},
\en
which agrees with the case of periodic boundary condition \cite{BAXTER}. Therefore, even in the situation where the boundary matrix becomes singular the physical quantities remain the same.

\section{Conclusion}
\label{CONCLUSION}

In this paper we have studied the dependence of physical quantities, like free-energy and entropy, of the six-vertex model on boundary conditions. 

We analyzed the case of free, periodic, anti-periodic and the mixture of periodic and anti-periodic boundary conditions for finite system sizes and in the thermodynamic limit. We have argued that all these boundary conditions produce the same results as the periodic boundary conditions.

We have also addressed the case of fixed boundary conditions and a variety of cases arise:

1. We found a family of boundary conditions for which the entropy coincide with its domain wall boundary conditions value at the ice-point. We called this boundary as descendent of the domain boundary conditions, once it has the same number of states as the domain wall boundary conditions and is simply related. We have provided a determinant solution for its partition function, whose homogeneous limit at the ice-point produce the same number of states and therefore the same entropy.

2. We have also found boundary conditions which are the mixture of domain wall and ferroelectric boundary conditions. This is obtained by the merge of the domain wall and the ferroelectric boundary conditions. There is also an additional possibility of the merge of two domain wall boundary conditions.

3. Besides that, we introduce what we called the N\'eel boundary condition. Due to the direct computation of the number of states, we argued that its entropy is the same as the periodic boundary conditions. 

4. The entropy per lattice site of the six-vertex model assumes the values $S_{FE}=0$ for ferroelectric boundary conditions and $S_{PBC}=\frac{1}{2} \ln\left(\frac{4}{3}\right)^3$ for periodic boundary conditions. Besides, the entropy $S_{DWBC}=\frac{1}{2} \ln\left(\frac{3^3}{2^4}\right)$ for the domain wall boundary condition is in between these two values. We have showed, for the case of the merge of domain wall and ferroelectric boundary condition, that the entropy vary from its values for the ferroelectric case to the domain wall boundary case. In addition, we propose that the merge of domain wall boundary with N\'eel or free boundary could result in entropy values in the interval between $S_{DWBC}$ and $S_{PBC}$.  Nevertheless, the explicit computation of the entropy in these cases is still missing. However, it is clear here that the entropy varies continously from $S_{FE}$ to  $S_{PBC}$.

In addition to that, we have shown that the free-energy of the six-vertex model at the isotropic point does not change even in the case that the transfer matrix becomes defective. In the case of singular toroidal boundary, the transfer matrix has fewer eigenstates than expected, however the free-energy in the thermodynamic limit remains the same.

Although we have found additional interesting fixed boundary conditions which produce entropies different from the periodic case, the complete classification of the boundary conditions in groups of similar pattern is still missing. Another interesting question would be if there exist other vertex models which depend on the boundary conditions. We hope to address to these problems in the future.

\section*{Acknowledgments}
T.S. Tavares thanks the C.N. Yang Institute for Theoretical Physics for hospitality and FAPESP for financial support through the grant 2013/17338-4. G.A.P. Ribeiro acknowledges financial support through the grant 2012/24514-0, S\~ao Paulo Research Foundation (FAPESP). V.E. Korepin was supported by NSF Grant DMS 1205422 and thanks P.M. Bleher for discussions.

\section*{\bf Appendix A: The eight-vertex model}\label{8vsec}
\setcounter{equation}{0}
\renewcommand{\theequation}{A.\arabic{equation}}

Here we summarize the results for the eight-vertex model.

The matrix elements of (\ref{countmixed}) for the eight-vertex model at the infinite temperature point ($a=b=c=d=1$) result in the number of states for all mixed boundary conditions. These numbers of physical states ($\Omega$) for the cases $N=L=2,3$ are given below,
\bear
M_{2,2}=2^5 \left(\begin{array}{cc|cc}
1 &  0 & 0 & 1 \\
0 &  1 & 1 & 0 \\
\hline
0 &  1 & 1 & 0 \\
1 &  0 & 0 & 1 \\
\end{array}\right),
M_{3,3}=2^{10}\left(\begin{array}{cccc|cccc}
1 &  0 & 0 & 1 & 0 & 1 & 1 & 0 \\
0 &  1 & 1 & 0 & 1 & 0 & 0  & 1 \\
0 &  1 & 1 & 0 & 1 & 0 & 0  & 1 \\
1 &  0 & 0 & 1 & 0 & 1 & 1 & 0 \\
\hline
0 & 1 & 1 & 0 & 1 & 0 & 0 & 1 \\
1& 0 &  0 & 1 & 0 & 1 & 1 & 0 \\
1& 0 &  0 & 1 & 0 & 1 & 1 & 0 \\
0 & 1 & 1 & 0 & 1 & 0 & 0 & 1
\end{array}\right).
\ear
where the non-zero elements satisfies the selection rule $\mbox{mod}\left[\Phi-\Theta,2\right]$. In the general case, one has that $\Omega_{N}=2^{N^2+1}$ for periodic, anti-periodic and mixed boundary conditions. The cases of fixed boundary conditions, e.g the  N\'eel boundary and domain wall boundary condition have the same number of states $\Omega_N^{DWBC}=\Omega_N^{NE}=2^{(N-1)^2}$ states. In any case, the entropy is in the thermodynamic limit $S=S_{DWBC}=S_{NE}=\ln{2}$. This confirm the independence of the eight-vertex model with the boundary condition\cite{WU}.

\section*{\bf Appendix B: Bethe ansatz solution for the leading contribution}\label{BAsol}
\setcounter{equation}{0}
\renewcommand{\theequation}{B.\arabic{equation}}

In order to show that the largest eigenvalue of different spin-$z$ components produce the same free-energy in the thermodynamic limit, we analyze the solution of the Bethe ansatz equation (\ref{Bet}).
We can assume $a>b+c$ in the ferroelectric phase. A possible parametrization for $\Delta > 1$ is
\begin{align}
a(\lambda)&=\varrho \sinh(\lambda+\gamma),\nonumber \\
b(\lambda)&=\varrho \sinh(\lambda), ~~ \Rightarrow~~  \Delta= \cosh(\gamma) ,\\
c(\lambda)&=\varrho \sinh(\gamma),\nonumber
\end{align}
where $\varrho$ is scale factor and $\gamma$ the anisotropy parameter.

We replace the above Boltzmann weights in the Bethe ansatz equations (\ref{Bet}) and its eigenvalue expression, where we also perform the shifts $\lambda\rightarrow\lambda-\gamma/2$ and $\lambda_i\rightarrow\lambda_i-\gamma/2$ for convenience.
Given $a,~b,~c$ in the region $a>b+c$, one can always obtain the corresponding $\gamma,~\varrho,~\lambda$. We checked the Bethe ansatz solution against the direct diagonalization in sector $n=L/2$ in order to identify the Bethe root structure leading to the largest eigenvalue of this sector. We have done this for $L$ up to $10$. We list the Bethe ansatz roots in the Table \ref{BAroots}, where we have chosen $a=2.1~, b=0.7,~c=0.62$. This choice for the Boltzmann weights corresponds to $\gamma\approx 0.93881$. Hence the above root structure resembles that of a $n=\frac{L}{2}$ string centered at $- \im \pi/2$ and lying along an axis parallel to the real axis \cite{GAUDIN,TAKAHASHI}. In the present case, this comes from the fact that whenever $\lambda_i$ in the left hand side of (\ref{Bet}) has some non-zero real part, then the ratio $a(\lambda_i-\gamma/2)/b(\lambda_i-\gamma/2)$ becomes either less than unity or greater than unity. Hence for large $L$ the left hand side tends to be zero or infinity. In order to 
have the same on the right hand side of (\ref{Bet}), one should have $\lambda_j-\lambda_i \rightarrow \pm \gamma$ for some $i$.
\begin{table}
\begin{center}
\begin{tabular}{|c|c|c|}
 \hline
$L$ & $\Lambda$  & Bethe roots \\ 
 \hline
$\small 2$ & $\small 3.3244$ &
\begin{tabular}{l}
$\small \lambda_{1}= 0. -1.5708 \im$
\end{tabular} \\ 
\hline
$\small 4$ & $7.15616$ &
\begin{tabular}{l}
 $\small\lambda_{1}= 0.755792 -1.5708 \im$ \\
 $\small\lambda_{2}= -0.755792+1.5708 \im$
\end{tabular}
 \\ \hline
 \small 6 & \small 18.9897 &
\begin{tabular}{l}
 $\small \lambda_{1}= 1.24738 -1.5708 \im$ \\
 $\small \lambda_{2}= 0. -1.5708 \im$ \\
 $\small\lambda_{3}= -1.24738-1.5708 \im$ 
\end{tabular}
 \\ \hline
$\small 8$ & $\small 59.0169$ &
\begin{tabular}{l}
 $\small\lambda_{1}= 1.67202 -1.5708 \im $\\
 $\small\lambda_{2}= 0.513821 -1.5708 \im $\\
 $\small\lambda_{3}= -0.513821-1.5708 \im $ \\
 $\small\lambda_{4}= -1.67202-1.5708 \im $
 \end{tabular}
 \\ \hline
 $\small 10$ & $\small 214.268$ &
\begin{tabular}{l}
$\small \lambda_{1}= 2.11256 -1.5708 \im $\\
$\small \lambda_{2}= 0.998365 -1.5708 \im$ \\
$\small \lambda_{3}= 0. -1.5708 \im $\\
$\small \lambda_{4}= -0.998365-1.5708 \im $\\
$\small \lambda_{5}= -2.11256-1.5708 \im $
\end{tabular}
\\\hline
\end{tabular}
\caption{Bethe ansatz roots for $a=2.1~, b=0.7,~c=0.62$.}
\label{BAroots}
\end{center}
\end{table}

\begin{figure}
  \begin{center}
  \includegraphics[width= 0.6 \linewidth]{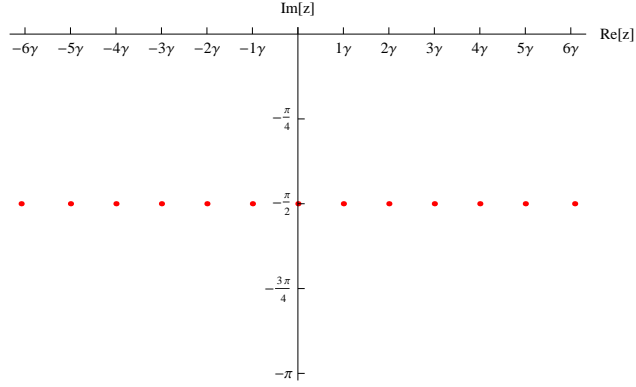}\\
  \caption{Distribution of Bethe roots leading to the largest eigenvalue in sector $L/2$ for $L=26$. There is small deviation from the $L/2$ string pattern.}\label{fig-string}
\end{center}
\end{figure}

In fact, the solution for the Bethe equations for $L=26$ has very small deviation from this $L/2$-string pattern, see Figure \ref{fig-string}.

Assuming that the exact string is given by
\eq
\lambda_j= -\im \frac{\pi }{2}+\gamma \left(\frac{L/2-1}{2}-(j-1)\right), \qquad j=1, \dots,L/2,
\en
we obtain a reasonable approximation for $\Lambda_{n=L/2}$ for large $L$ as follows,
\bear
\frac{\Lambda_{\max,n=L/2}(\lambda)}{\varrho^L} &\approx& \sinh^L(\lambda+\gamma/2) \prod_{q=-L/4+1/2}^{L/4-1/2} \frac{\cosh(\lambda+q\gamma -\gamma)}{\cosh(\lambda+q \gamma)} \nonumber \\
&+&\sinh^L(\lambda-\gamma/2) \prod_{q=-L/4+1/2}^{L/4-1/2} \frac{\cosh(\lambda+ q \gamma+\gamma)}{\cosh(\lambda+q \gamma)}, \\
&=& \sinh^L(\lambda+\gamma/2) \frac{\cosh(\gamma (L/4-1/2)\gamma-\lambda)}{\cosh(\gamma (L/4-1/2)+\lambda)} \nonumber \\
&+& \sinh^L(\lambda-\gamma/2) \frac{\cosh(\gamma (L/4-1/2)+\gamma+\lambda)}{\cosh(\gamma (L/4-1/2)-\lambda)},\\
&\approx& \sinh^L(\lambda+\gamma/2) {\rm e}^{\gamma-2 \lambda}+\sinh^L(\lambda-\gamma/2) {\rm e}^{\gamma+ 2 \lambda}.
\ear
Taking the thermodynamic limit we find
\eq
e^{-\beta F_{PA}}= \lim_{L \rightarrow \infty} {\left(\Lambda_{\max,n=L/2}(\lambda)\right)}^{\frac{1}{L}}=  a(\lambda).
\en
This is the same free-energy as the case of periodic boundary in the ferroelectric regime for $a>b$ \cite{BAXTER}.

\section*{\bf Appendix C: Separated inversions}\label{appC}
\setcounter{equation}{0}
\renewcommand{\theequation}{C.\arabic{equation}}

In the case of separated inversions we have
\begin{multline}
Z_{j,1}= \sum_{g_1,g_2',g_3,g_4'}{\left(\Lambda_{g_1}^{(0)}\right)}^{N-m_2+m_1-1} {\left(\Lambda_{g_2}^{(1)}\right)} {\left(\Lambda_{g_3}^{(0)}\right)}^{m_2-m_1-1} {\left(\Lambda_{g_4}^{(1)}\right)} \times\\ \times \braked{g_1^{(0)}\big \vert g_2^{(1)}} \braked{g_2^{(1)}\big \vert g_3^{(0)}} \braked{g_3^{(0)}\big \vert g_4^{(1)}} \braked{g_4^{(1)}\big \vert g_1^{(0)}} {\left[1+{(-1)}^{1+\alpha_{g_1}+\alpha_{g_3}}\right]}^2,
\end{multline}
therefore $\alpha_{g_1}$ and $\alpha_{g_3}$ must be different for nonzero contributions. Hence in the above summation we should restrict $\{g_1,g_3\}$ to the sequences where $\alpha_{g_1}=1-\alpha_{g_3}$.

For $L$ odd, sectors $n$ and $L-n$ have different parities. Besides, as we can read from relations (\ref{symT0}) and (\ref{symPXPZ}), for each eigenvector $\ket{g}$ in sector $n$ we have another eigenvector $\Pi^x \ket{g}$ in sector $L-n$ with the same eigenvalue. Therefore, for $L$ odd we obtain
\eq
F_{j,1}= \lim_{L,N \rightarrow \infty} -\frac{1}{\beta N L} \ln \left(4 \sum_{n=0}^L {\left(\Lambda_{\max, n}^{(0)}\right)}^{N-2} {\left(\Lambda_{\max}^{(1)}\right)}^2 {\left \vert \braked{g_{\max,n}^{(0)} \big \vert g_{\max}^{(1)}} \right \vert}^4 \right)=F_{PP}.
\en
under assumption of (\ref{hip1}).

For $L$ even the sectors $n$ and $L-n$ have the same parity. In this case, a possible approximation  is to retain the maximum eigenvector in sector $n$ for $g_1$ and use the maximum eigenvector in sectors $n \pm 1$ and $L-n \pm 1$ for $g_3$. Therefore we find
\bear
F_{j,1}&=& -\lim_{L,N \rightarrow \infty} \frac{1}{\beta N L} \ln \Bigg( 8 \sum_{n=0}^L\Bigg[\sum_{|m-n|=1} {\left(\Lambda_{\max,n}^{(0)}\right)}^{N-m_2+m_1-1} {\left(\Lambda_{\max,m}^{(0)}\right)}^{m_2-m_1-1}  \nonumber \\ 
&\times& \left(\Lambda_{\max}^{(1)}\right)^{2} {\left \vert \braked{g_{\max,n}^{(0)} \big \vert g_{\max}^{(1)}} \braked{g_{\max,m}^{(0)} \big \vert g_{\max}^{(1)}} \right \vert}^2 \Bigg] \Bigg)=F_{PP},
\ear
%% tem m_1 e m_2 na soma, arrumar isto...
where we are making use of hypothesis (\ref{hip1}) and also the following:
\eq
\lim_{L \rightarrow \infty}{\left(\Lambda_{\max,n}^{(0)}\right)}^{\frac{1}{L}}=\lim_{L \rightarrow \infty}{\left(\Lambda_{\max,n\pm 1}^{(0)}\right)}^{\frac{1}{L}}.
\en

\end{document}